\documentclass[prx, reprint, superscriptaddress, amsmath, amssymb, aps]{revtex4-2}

\usepackage{graphicx}
\usepackage{dcolumn}
\usepackage{bm}
\usepackage[protrusion=true,expansion=true]{microtype}
\usepackage{hyperref}
\usepackage{cleveref}

\def\adhoc{{\bf ad hoc }}
\def\bfadhoc{{\textbf{ad hoc}}}
\def\min{\textrm{min}}
\def\max{\textrm{max}}

\def\cm{\textrm{cm}}
\def\ti{\textrm{i}}

\def\bara{\bar a}
\def\barp{\bar p}
\def\bars{\bar s}
\def\barh{\bar h}

\DeclareMathOperator{\sech}{sech}

\begin{document}

\preprint{APS/123-QED}

\title{Cluster-Mediated Synchronization Dynamics in \\ Globally Coupled Oscillators with Inertia} 


\author{Cook Hyun Kim}
\affiliation{CCSS, KI for Grid Modernization, Korea Institute of Energy Technology, 21 Kentech, Naju, Jeonnam 58330, Korea}

\author{Jinha Park}%
\affiliation{CCSS, KI for Grid Modernization, Korea Institute of Energy Technology, 21 Kentech, Naju, Jeonnam 58330, Korea}

\author{Young Jin Kim}%
\affiliation{Division of Data Analysis, Korea Institute of Science and Technology Information, 66 Hoegiro, Seoul 024256, Korea}

\author{Sangjoon Park}%
\affiliation{CCSS, KI for Grid Modernization, Korea Institute of Energy Technology, 21 Kentech, Naju, Jeonnam 58330, Korea}

\author{S. Boccaletti}%
\affiliation{Sine-Europe Complexity Science Center, North University of China, 3 Xueyuan, Taiyuan, Shanxi, 030051, China}
\affiliation{Research Institute of Interdisciplinary Intelligent Science, Ningbo University of Technology, 201 Fenghua, Ningbo, Zhejiang, 315211, China}
\affiliation{CNR, Institute of Complex Systems, Madonna del Piano 10, Sesto Fiorentino, Firenze, 50019, Italy}

\author{B. Kahng}
\email{bkahng@kentech.ac.kr}
\affiliation{CCSS, KI for Grid Modernization, Korea Institute of Energy Technology, 21 Kentech, Naju, Jeonnam 58330, Korea}

\date{\today}

\begin{abstract}
Globally coupled oscillator systems with inertia exhibit complex synchronization patterns, among which the emergence of a couple of secondary synchronized clusters (SCs) in addition to the primary cluster (PC) is especially distinctive. Although previous studies have predominantly focused on the collective properties of the PC, the dynamics of individual clusters and their inter-cluster interactions remain largely unexplored. Here, we demonstrate that multiple clusters emerge and coexist, forming a hierarchical pattern known as the Devil's Staircase. We identify three key findings by investigating individual cluster dynamics and inter-cluster interactions. First, the PC persistently suppresses the formation of SCs during its growth and even after it has fully formed, revealing the significant impact of inter-cluster interactions on cluster formation. Second, once established, SCs induce higher-order clusters exhibiting frequency resonance via inter-cluster interactions, resulting in the Devil's Staircase pattern. Third, sufficiently large SCs can destabilize and fragment the PC, highlighting the bidirectional nature of cluster interactions. We develop a coarse-grained Kuramoto model that treats each cluster as a macroscopic oscillator to capture these inter-cluster dynamics and the resulting phenomena. Our work marks a significant step beyond system-wide averages in the study of inertial oscillator systems, offering new insights into the rich dynamics of cluster formation and synchronization in real-world applications such as power grid networks.
\end{abstract}

\maketitle

\section{Introduction}\label{sec:Introduction}

Synchronization is a ubiquitous collective behavior observed in various natural and engineered systems, from neural networks~\cite{bick2020understanding,cumin2007generalising,breakspear2010generative} and power grids~\cite{grainger1999power, anvari2020introduction, witthaut2022collective, menck2013basin} to chemical reactions~\cite{sivashinsky1977diffusional,forrester2015arrays} and biological rhythms~\cite{buck1938synchronous, buck1966biology, buck1976synchronous, buck1988synchronous}. The Kuramoto model, introduced in 1975, has become a paradigmatic framework for understanding such synchronization phenomena~\cite{kuramoto1975international,kuramoto1984chemical}. While the first-order Kuramoto model has provided substantial insights into coupled oscillator dynamics, many real-world systems exhibit more complex behaviors that necessitate consideration of inertial effects, leading to the development of the second-order Kuramoto model~\cite{strogatz2000kuramoto,acebron2005kuramoto}, which reveals richer dynamics including hysteresis and discontinuous transitions~\cite{tanaka1997first,tanaka1997self,olmi2014hysteretic,gao2018self,gao2021synchronized}.

Previous studies have primarily focused on system-wide averages, particularly macroscopic order parameters, to characterize the system's behavior. While this system-wide approach has been valuable for understanding global synchronization, it has left the dynamical properties of multiple clusters - specifically, their mutual interactions and cluster-level dynamics - largely unexplored. Understanding these dynamical properties of multiple clusters is essential not only for theoretical insights but also for practical applications in stable and efficient power grid operation~\cite{grainger1999power,anvari2020introduction} and neural synchronization~\cite{bick2020understanding}.

Here, we develop a coarse-grained Kuramoto model that treats each cluster as a single macroscopic oscillator, effectively reducing the high-dimensional dynamics of individual oscillators to a lower-dimensional system of clusters. This dimensional reduction not only simplifies the equation of motion but also allows us to identify and characterize the interactions among clusters, providing a framework for understanding the dynamics of multiple frequency-synchronized clusters. Our approach uncovers three distinct phenomena in inter-cluster interactions: First, we demonstrate the formation of a primary cluster (PC) through an \adhoc potential. During and after the PC’s growth, the PC suppresses the formation of secondary clusters (SCs) through differential variations in potential energy across individual oscillators that generate repulsive interactions, thereby impeding frequency synchronization. Second, once the PC stabilizes, SCs and higher-order clusters (HOCs) arise near their orbital aphelion, where their potential energy is maximized and kinetic energy is minimized. This leads to extended dwelling times and enhanced interactions among clusters, resulting in a hierarchical pattern with rational frequency ratios—a Devil’s Staircase. Third, if SCs grow sufficiently large, the PC can destabilize and collapse under the competing influences of the SCs’ attractive forces and the PC’s internal cohesion. This collapse manifests as a sudden disintegration of the PC, with its oscillators either merging completely into SCs or oscillating coherently between them while preserving time-averaged frequency synchronization.

Our results advance the theoretical understanding of synchronization in systems with inertia and provide practical insights for applications ranging from power grid management to neural network design~\cite{guo2021overviews}. The cluster-level approach presented here opens new avenues for investigating synchronized cluster dynamics in various physical and biological systems where multiple clusters demonstrate complex and diverse dynamical behaviors.


\begin{figure*}[!]
\resizebox{1.0\linewidth}{!}{\includegraphics{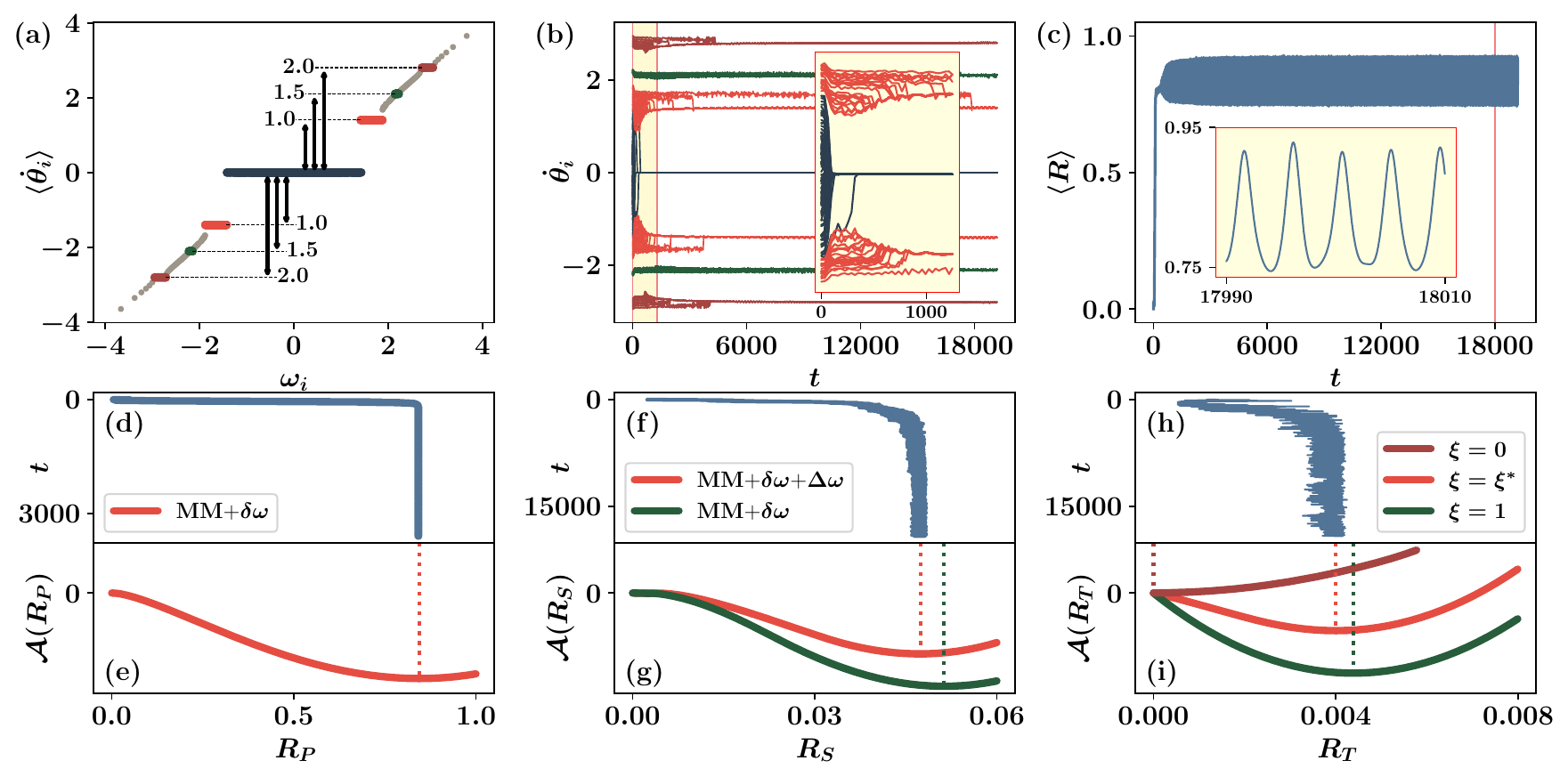}}
\caption{\textbf{Cluster Synchronization and \bfadhoc Potential}
(a) Steady-state mean angular velocity $\langle \dot \theta_i \rangle$ versus natural frequency $\omega_i$, showing a Devil's Staircase pattern with distinct plateaus: PC (dark blue, $\langle \dot \theta_i \rangle \simeq 0$), SC (bright red), TC (dark green), and QC (deep red). The relative frequencies of SC, TC, and QC to PC are given by the rational ratios 1:1.5:2. (b) Angular velocity evolution $\dot \theta_i(t)$ for oscillators in different clusters, using the same color scheme as in (a). The inset shows how these clusters form sequentially after the onset of the PC cluster. (c) The evolution of order parameter $R$ illustrates the PC cluster's initial development followed by the SC cluster's emergence. (d)--(i) Evolution of order parameters and corresponding \adhoc\ potentials: (d,f,h) Time series of the order parameters $R_P$, $R_S$, and $R_T$, respectively, plotted with time increasing from top to bottom along the $y$-axis. These data were obtained using the fourth-order Runge–Kutta (RK-4) method. (e,g,i) Shows the corresponding \adhoc\ potentials as functions of the respective order parameters, calculated with specific boundaries for PC, SC, and TC. The bright red, dark green and deep red curves represent different boundary formulations (see the main text for details). All calculations were performed with $(m,K)=(8,12)$.}
\label{fig:fig1}	
\end{figure*}


\section{Main Results}
\subsection{Coarse-Grained Equation of Motion}

To systematically analyze cluster formation and dynamics, we begin with the second-order Kuramoto model~\cite{kuramoto1975international,kuramoto1984chemical,strogatz2000kuramoto,acebron2005kuramoto}, which describes an ensemble of $N$ globally coupled phase oscillators:
\begin{align}
m \ddot{\theta}_{i} + \gamma \dot{\theta}_{i} = \omega_{i} + \dfrac{K}{N}\sum_{\ell=1}^{N} \sin(\theta_{\ell} - \theta_{i}) \equiv \omega_{i} + \tau_{i}.
\label{eq:Kuramoto_Model}
\end{align}
Here, $\theta_{i}$ and $\dot{\theta}_{i}$ represent the phase and angular velocity of oscillator $i$ ($i=1,\cdots, N$), $\omega_i$ is its intrinsic frequency drawn from a unimodal distribution $g(\omega)$, and $m$ and $\gamma$ are the inertial mass and dissipation coefficient, respectively. The coupling strength $K$ determines how strongly oscillators interact, with $\tau_i$ representing the total torque exerted on oscillator $i$ by all other oscillators. When $K$ exceeds a critical threshold $K_c$, the system transitions from disorder to synchronized states.

The key to understanding cluster formation lies in decomposing the total torque into contributions from different clusters:
\begin{align}
\tau_{i} =\tau_{p,i} + \tau_{s,i} + \tau_{h,i} + \tau_{\circ,i},
\label{eq:torque_decomp}
\end{align}
where
\begin{align}
\tau_{a,i}={\frac{K}{N}}\sum_{j\in C_a}\sin(\theta_{j}-\theta_{i}), \, \tau_{\circ,i} = {\dfrac{K}{N}} \sin(\theta_{\circ} - \theta_{i}).
\label{eq:torque_a_i}
\end{align}
In this decomposition, $\tau_{a,i}$ represents the torque exerted on oscillator $i$ by all oscillators in cluster $a$, where $a$ can be primary (p), secondary (s), or higher-order (h) clusters. The term $\tau_{\circ,i}$ accounts for oscillators not belonging to any specific cluster.

This cluster-based perspective allows us to introduce a powerful analytical tool: the concept of a virtual super-oscillator representing each cluster's collective behavior. For cluster $a$, this super-oscillator has mean phase $\theta_{\bar a}$ and mean angular velocity $\dot \theta_{\bar a}$. In the steady state, individual oscillators maintain fixed relationships with their cluster's super-oscillator:
\begin{align}
\langle \theta_{i}-\theta_{\bar a} \rangle_{t} \simeq \text{constant}, \quad \langle \dot{\theta}_{i}-\dot{\theta}_{\bar a} \rangle_{t} \simeq 0,
\label{eq:phase_freuquency_T-AV}
\end{align}
where $\langle \cdots \rangle_t$ denotes time averaging.
Using this super-oscillator framework, we can express the torque exerted on oscillator $i\in C_a$ by all oscillators in its cluster $C_a$ as:
\begin{align}
\tau_{a,i} = K R_{a} \sin \left(\theta_{\bar a} - \theta_i\right),
\end{align}
where $R_{a}\exp(\ti\theta_{\bar a})= (1/N) \sum_{j \in C_a} \exp(\ti \theta_{j})$ represents the complex order parameter for cluster $a$. This elegant formulation captures both the collective behavior of the cluster through $R_a$ and the individual oscillator's deviation through the phase difference $\theta_{\bar a} - \theta_i$.

With this foundation, we can analyze the relative motion of oscillators to any cluster's center of mass. The governing equation (the coarse-grained second-order Kuramoto model) for oscillator $i\in C_a$ becomes:
\begin{align}
m \ddot{\theta}_{i,\bar a} + \gamma \dot{\theta}_{i, \bar a}
= \omega_{i}-\omega_{\bar a}+\tau_{p,i}-\tau_{p,{\bar a}} + \tau_{s,i} - \tau_{s,\bar a}.
\label{eq:relative_angle}
\end{align}
Here, $\theta_{i,\bar a} =\theta_{i}-\theta_{\bar a}$ describes the phase relative to the cluster's mean, and $\omega_{\bara} = \langle \omega_{i} \rangle_{i \in C_a}$ is the cluster's coarse-grained intrinsic frequency. In deriving this equation, we have neglected torques from HOCs and unaffiliated oscillators due to their relatively small magnitude. This approximation lets us focus on the essential dynamics governing cluster formation and stability.

\subsection{Primary Cluster}

A PC forms the foundation of the system's synchronization phenomena. For this cluster ($a=p$), Eq.~\eqref{eq:relative_angle} takes a particularly simple form:
\begin{align}
m \ddot{\theta}_{i,\bar p} + \gamma \dot{\theta}_{i, \bar p} \simeq \omega_{i} - \omega_{\bar p} + \tau_{p,i},
\label{eq:E-M_p}
\end{align}
where other torque terms can be neglected due to the PC's dominance in the early stage (PC's formation stage) of synchronization. This equation describes how oscillators in the PC interact exclusively within their cluster, highlighting the PC's self-contained nature during initial formation. 

When oscillators are bound to the PC, the oscillators reach a fixed point where inertial and dissipative terms vanish, leading to a torque balance equation:
\begin{align}
\omega_{i}-\omega_{\bar p}+KR_{P}\sin(\theta_{\bar p}- \theta_{i})= 0.
\label{eq:omega_ip_pc}
\end{align}
This balance determines the phase angles of the oscillators relative to the center of mass within their cluster. Based on this torque balance condition, the PC's formation and stability are governed by a self-consistency equation for its order parameter:
\begin{align}
R_P =
\int_{\omega_{p,\min}}^{\omega_{p,\max}} \sqrt{1-\left(\dfrac{\omega - \omega_{\bar p}}{K R_{P}}\right)^2} g(\omega) d\omega
\equiv f(R_P).
\label{eq:sce_p}
\end{align}
The integration limits define the PC's frequency boundaries, which we determine using the Melnikov method:
\begin{align}
\label{eq:omega_p_range}
\omega_{p,\max} & = \omega_{\bar p} + \left({4 \gamma}/{\pi}\right) \sqrt{{K R_{P}}/{m}} \mathrel{+} \delta\omega_{p},\cr
\omega_{p,\min} & = \omega_{\bar p} - \underbrace{\left({4 \gamma}/{\pi}\right) \sqrt{{K R_{P}}/{m}}}_{\rm MM} \mathrel{-} \delta\omega_{p},
\end{align}
where $\delta\omega_{a} \simeq \left({\alpha}/{m}\right)\sqrt{1/(KR_{a}{m})}$ represents a correction term. This correction, originally introduced for homoclinic bifurcation analysis~\cite{belykh2016bistability}, proves essential for accurately predicting PC's formation in the second-order Kuramoto model~\cite{tanaka1997first, tanaka1997self, olmi2014hysteretic, gao2018self, gao2021synchronized}. The empirical value $\alpha=0.3056$ provides remarkable agreement with numerical observations. Note that the frequency scales follow $O(\omega)=O(KR_P)= O(1/m)$, reflecting the fundamental relationship between inertia and synchronization boundaries.

To visualize and analyze the PC's collective behavior in parameter space, we introduce an \adhoc potential $\mathcal{A}(R_P)$~\cite{song2020effective, jhun2022quantum} derived from self-consistency equation:
\begin{align}
\mathcal{A}(R_{P})\equiv K\int_{0}^{R_{P}} \big[R_{P}^\prime - f({R}_{p}^\prime)\big]d{R}_{p}^\prime,
\label{eq:adhoc}
\end{align}
where $\mathcal{A}(0)=0$ serves as a reference point. This potential provides key insights into the PC's stability: its minima, determined by $\partial \mathcal{A}/\partial R_P\big|_{R_P=R_P^*}=0$, correspond to stable solutions $R_P^*$ of the self-consistency equation. As shown in [Fig.~\ref{fig:fig1}(e)], these analytical predictions align remarkably well with the steady-state solutions $R_P(\infty)$ obtained through numerical Runge-Kutta integration [Fig.~\ref{fig:fig1}(d)]. This agreement validates our boundary predictions from Eq.~\eqref{eq:omega_p_range} and confirms the theoretical framework's accuracy in describing PC's formation and stability. For detailed derivations of the Melnikov method used in determining PC boundaries and further analysis of the \adhoc potential under various parameter conditions, see Supplementary, Sec.~\ref{sec:The Melnikov method: PC boundary} and~\ref{sec:adhoc potential of PC}.

\begin{figure*}
\resizebox{1.0\linewidth}{!}{\includegraphics{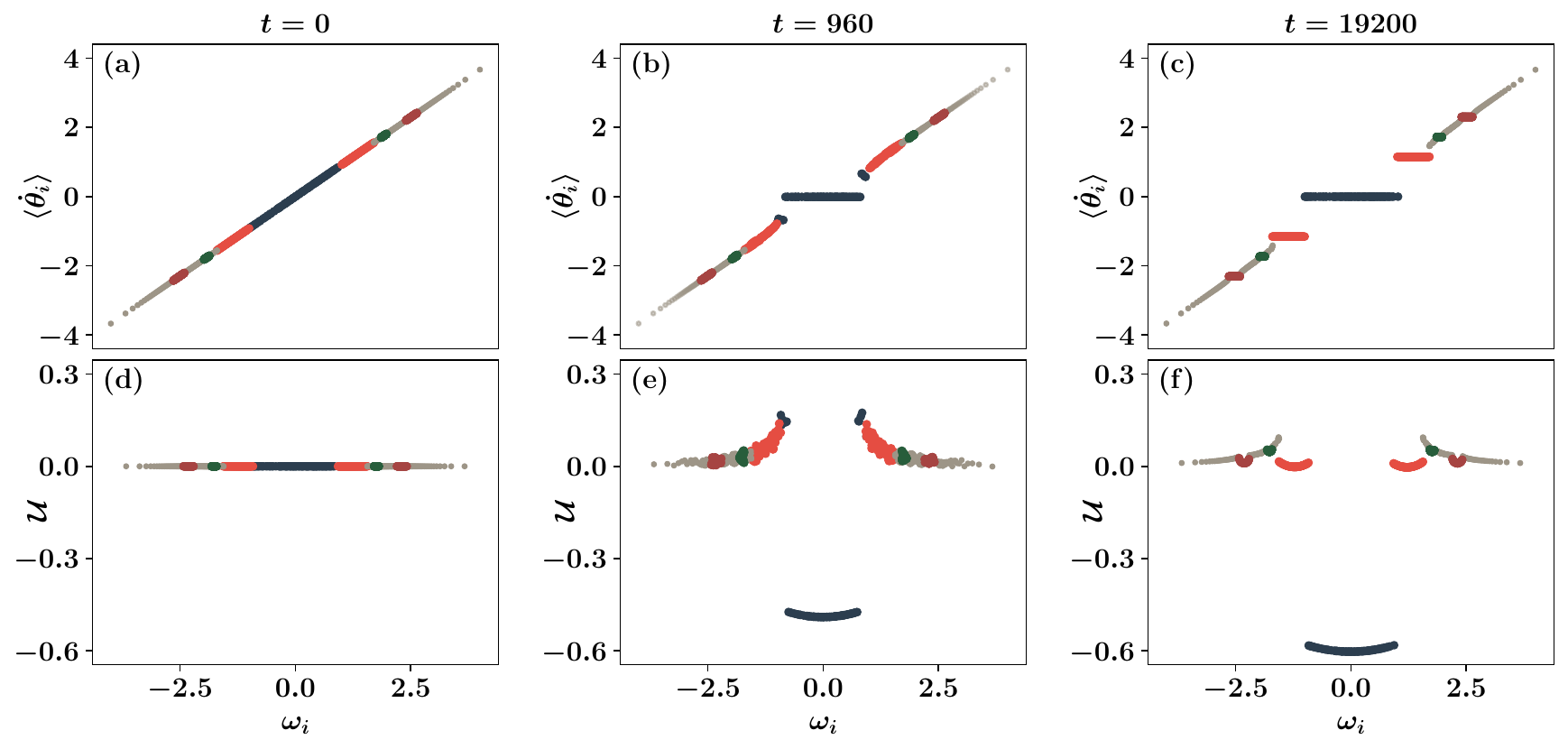}}
\centering
\caption{
\textbf{Evolution of average angular velocities and potential energy landscapes.} Upper row: Snapshots of mean angular velocity $\langle \dot \theta_i \rangle$ as a function of $\omega_i$ at different time points [(a) $t=0$, (b) $t=960$, and (c) $t=19200$], where the average is taken over a time window (b) $\pm 16$ and (c) $\pm 1024$ centered at each point. Lower row: Potential energy $\mathcal{U}$ as a function of $\omega_i$ at (d) $t=0$, (e) $t=960$, and (f) $t=19200$. The system develops a potential well for the PC at (e) $t=960$, eventually forming global and local wells corresponding to PC (dark blue), SC (bright red), TC (dark green), and QC (deep red) at (f) $t=19200$. All panels adopt the parameter values $(m,K)=(8,6)$.
}
\label{fig:fig3}	
\end{figure*}

\subsection{Potential Well}

The formation of SCs and HOCs follows a temporal hierarchy, emerging only after the PC has fully developed. This sequence becomes apparent when examining the angular velocity $\dot{\theta}_{i}$ over time [Fig.~\ref{fig:fig1}(b)]. To understand this behavior in cluster level, we analyze the evolution of each oscillator's potential energy:
\begin{align}
\mathcal{U}(\theta_i(t))= -
\sum_{a \in (p,s,h)} KR_{a}\cos{(\theta_i(t) - \theta_{\bar a})},
\label{eq:potential_total}
\end{align}
This potential describes the interaction energy between each oscillator ($\theta_{i}$) and all existing clusters ($\theta_{\bar a}$).
The system's potential energy landscape evolves through several distinct stages [Fig.~\ref{fig:fig3}]. From an initially flat configuration, a primary potential well emerges, corresponding to the PC's formation. As this well deepens during the PC's growth, a decrease in the PC's potential energy induces a corresponding increase in the remaining oscillators' potential energy. This energy framework explains the initial suppression of additional cluster formation: the continuously increasing potential energies act as repulsive interactions among oscillators, preventing them from forming stable synchronized clusters.
The emergence of additional clusters becomes possible only after the PC reaches its minimum potential energy state. At this point, the PC well stabilizes, with no further decrease in its potential energy, thereby halting the increase in other oscillators' potential energy, which is equivalent to the absence of repulsive interactions among these oscillators. Only then do new potential wells develop, corresponding to the formation of SCs and HOCs. This sequence of well formation reveals the underlying mechanism of the system's hierarchical organization: the stabilization of each cluster's potential energy creates favorable conditions for subsequent cluster formation. For a detailed analysis of how PC suppresses the formation of SCs, see Supplementary, Sec.~\ref{sec:SC forms after PC growth}.

\subsection{Secondary Clusters}

The formation of SCs reveal a subtle interplay between internal organization and external influences. For oscillators in an SC, the coarse-grained equation takes the form:
\begin{align}
m \ddot{\theta}_{j,\bar s} + \gamma \dot{\theta}_{j, \bar s} \simeq \omega_{j} - \omega_{\bar s} + \tau_{s,j} + \tau_{p,j} - \tau_{p,{\bar s}}.
\label{eq:E-M_s}
\end{align}
This equation captures a crucial distinction from PC's formation: the first three terms ($\omega_{j}$, $\omega_{\bar{s}}$, and $\tau_{s,j}$) represent internal dynamics within the SC, while the remaining terms arise from interaction with the established PC.

The SC's evolution follows a characteristic pattern. The SC order parameter $R_S(t)$ grows continuously from zero to a steady-state value $R_S(\infty)$ [Fig.~\ref{fig:fig1}(e)]. Like the PC, the SC's frequency boundaries can be determined analytically. Our analysis yields:
\begin{align}
\label{eq:omega_s_range}
\omega_{s,\max} & = \omega_{\bar{s}} + \overbrace{\left({4 \gamma}/{\pi}\right) \sqrt{{K R_{S}}/{m}}}^{\rm MM} + \delta\omega_{s} + \Delta \omega_{s, \max}, \\
\omega_{s,\min} & = \max(\omega_{\bar{s}} - \left({4 \gamma}/{\pi}\right) \sqrt{{K R_{S}}/{m}} - \delta\omega_{s} - \Delta \omega_{s, \min}, \omega_{p,\max}), \nonumber
\end{align}
where the correction terms capture PC-SC interactions:
\begin{align}
\Delta \omega_{s, \max} & = \beta \dfrac{\gamma}{2} \left({K R_{P}}/{m}\right)^2\left\{{(\omega_{\bars}/\gamma)^{-3}} - {[(\omega_{\bars} + {\rm MM})/\gamma]^{-3}} \right\}, \cr
\Delta \omega_{s, \min} & = \beta \dfrac{\gamma}{2} \left({K R_{P}}/{m}\right)^2\left\{ {[(\omega_{\bars} - {\rm MM})/\gamma]^{-3}} - {(\omega_{\bars}/\gamma)^{-3}} \right\}.
\end{align}

These boundaries reveal several key features. First, two distinct SC regions emerge symmetrically around $\omega = 0$, reflecting the system's inherent symmetry. Second, the lower bound $\omega_{s,\min}$ must exceed $\omega_{p,\max}$, as oscillators within the PC's frequency range are already committed to that cluster. The parameter $\beta$, while independent of time, varies with system parameters $m$ and $K$ and is determined empirically to match observed dynamics [Fig.~\ref{fig:fig1}(f)]. For detailed derivations of how PC-SC interaction dynamics affects the Melnikov method in determining SC boundaries and further analysis of the \adhoc potential under various parameter conditions, see Supplementary, Sec.~\ref{sec:SC boundary} and~\ref{sec:adhoc potential for SC}.

The formation of SCs, however, is not guaranteed even within these boundaries. Consider two cases with identical $K/m$ ratios: $(m,K)=(3,4.8)$ and $(5,8)$. Despite sharing the same ratio, only the latter case supports SC formation [Figs.~\ref{fig:fig_S_8}(e) and (f)]. This reveals that SC formation depends critically on the balance between intrinsic frequency and clustering torque. The clustering tendency, determined by $\tau_{s,j}$ and $\omega_j-\omega_{\bar s}$, scales differently with $K$ and $m$: while the frequency range depends on $K/m$, the torque scales with $K$ alone. Thus, larger absolute values of $K$ facilitate clustering even at fixed $K/m$. For a detailed demonstration of the absence of SCs at small inertia, see Supplementary, Sec.~\ref{sec:when $m$ is small}.

\subsection{Higher Order (Tertiary, Quaternary) Clusters}

\begin{figure}[ht]
\centering
\resizebox{1.0\linewidth}{!}{\includegraphics{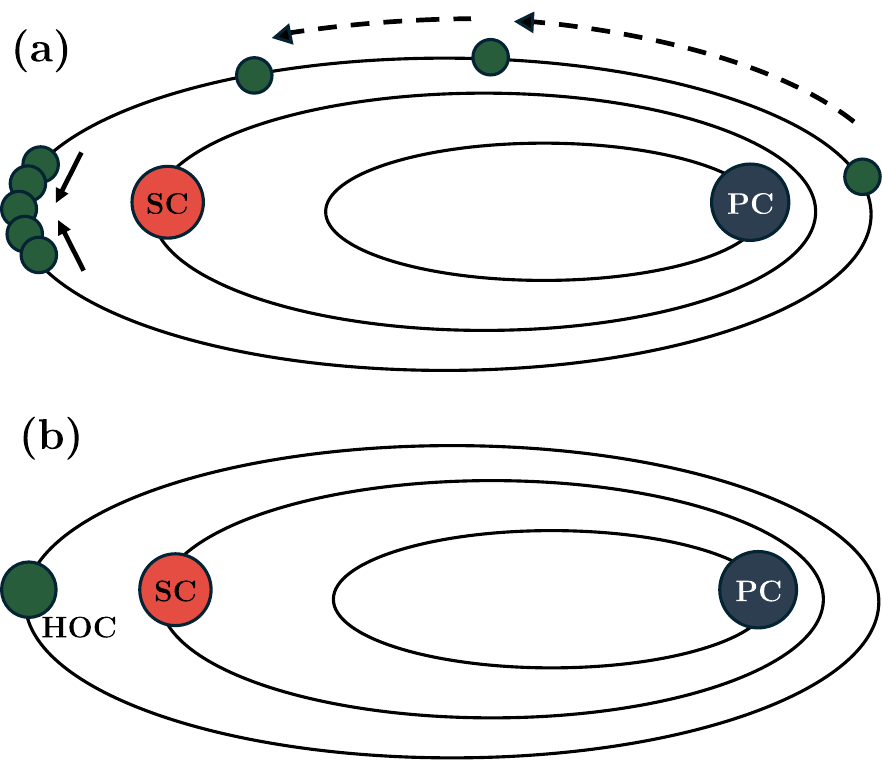}}
\caption{\textbf{Schematic illustration of HOC formation near the SC.} (a) The oscillators in SC and HOC exhibit orbital motion around the PC, with increased velocities near perihelion and decreased velocities near aphelion. (b) The reduced velocity near the aphelion enhances the interaction strength between SC and HOC oscillators, leading to cluster formation in this region. The SC establishes resonance with HOC oscillators, resulting in rotational frequencies that are rational multiples of the SC frequency.}
\label{fig:fig4}	
\end{figure}

We consider the formation of an HOC. Similar to the SC, a HOC forms near the aphelion of the PC, where the potential energy reaches its maximum. This maximum potential energy minimizes the kinetic energy of the oscillator, resulting in the smallest angular velocity at that point---a condition that strongly promotes inter-cluster interaction and cluster formation. The schematic illustration of these mechanisms is depicted in [Fig.~\ref{fig:fig4}]. Consequently, both SC and HOCs form in close proximity around the aphelion point, where their angular velocities become resonant. Specifically, the angular velocities of HOCs maintain rational-number ratios with respect to the SC frequency, as demonstrated in Supplementary, Sec.~\ref{sec:Devil's Staircase Pattern}. For instance, the averaged frequency ratios of TC and QC relative to SC are $\approx 1.5$ and $\approx 2.0$, respectively.


The resonance effect can be further analyzed using the \adhoc potential. Taking into account the SC's influence on HOC formation, we express the order parameter $f(R_H)$ empirically as
\begin{align}
f(R_H) = \sum_{k \in h} \sqrt{1 - \left(\dfrac{\omega_{k} - \omega_{\barh}}{K(\xi R_S+R_H)}\right)^2},
\label{eq:Work_Energy_H}
\end{align}
where $\omega_{\bar h}$ represents the mean frequency of HOC oscillators. Here, $K(\xi R_S+R_H)$ replaces $KR_P$ from Eq.~\eqref{eq:sce_p}, reflecting the combined influence of SC and HOC interactions. The parameter $\xi$, while independent of time, varies with system parameters $m$ and $K$, and is determined empirically to match observed dynamics [Fig.~\ref{fig:fig1}(h)]. For detailed analysis of the SC's influence on HOC formation and derivations of the \adhoc potential under various parameter conditions, see Supplementary, Sec.~\ref{sec:SC effect on the HOC} and~\ref{sec:adhoc potential of HOC}.

\begin{figure*}[!]
\resizebox{1.0\linewidth}{!}{\includegraphics{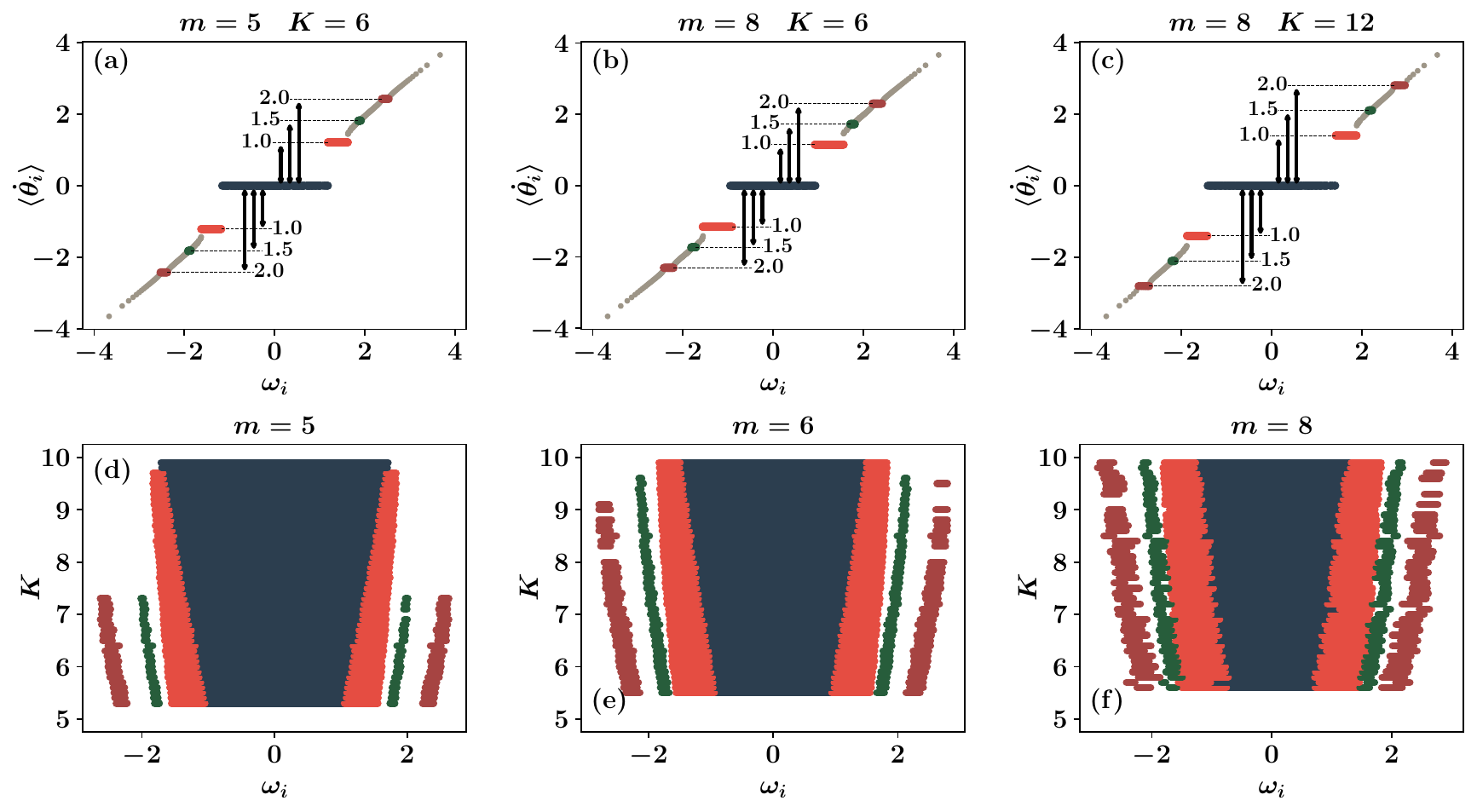}}
\centering
\caption{\textbf{Devil's Staircase patterns and Arnold's tongues}
(a)--(c) Mean angular velocity $\langle \dot \theta_i \rangle$ versus natural frequency $\omega_i$ showing the Devil's Staircase patterns for parameter sets: (a) $(m,K)=(5,6)$, (b) $(8,6)$, and (c) $(8,12)$. The data, computed from steady-state time series, reveal distinct clusters: PC (dark blue), SC (bright red), TC (dark green), and QC (deep red) with frequency gaps of 1, 3/2, and 2 from PC, respectively. The cluster sizes vary with $(m,K)$ values. (d)--(f) Corresponding Arnold's tongues showing oscillators belonging to different clusters in the $(K,\omega)$ parameter space, colored according to their cluster membership, for different inertia values: (d) $m=5.0$, (e) $m=6.0$, and (f) $m=8.0$. The stability regions of the Devil's Staircase patterns extend over a broader range of coupling strength $K$ with increasing inertia $m$.}
\label{fig:fig5}	
\end{figure*}

[Fig.~\ref{fig:fig5}(a)--(c)] illustrates the Devil's staircase patterns for various $(m,K)$ combinations. At fixed $K=6$, decreasing (increasing) $m$ widens (narrows) the PC range $\omega_p$ while narrowing (widening) the frequency ranges of non-primary clusters [Figs.~\ref{fig:fig5}(a) and (b)]. This indicates that reduced kinetic energy (smaller $m$) enhances the PC's clustering tendency, consequently diminishing the sizes of SCs and HOCs. Similarly, at fixed $m$, increasing $K$ strengthens the potential energy between oscillators, causing the PC range $\omega_p$ to widen while narrowing the frequency ranges of non-primary clusters [Figs.~\ref{fig:fig5}(b) and (c)].
This behavior manifests clearly in Arnold's tongue pattern shown in [Fig.~\ref{fig:fig5}(d)--(f)]. As $m$ decreases (increases), the plateau sizes of HOCs narrow (wide), with TCs and QCs failing to form at large $K$ values when $m$ is sufficiently small [Fig.~\ref{fig:fig5}(d)]. In contrast, at larger values of inertia (e.g., $m=8$), higher-order synchronization patterns emerge more distinctly in Arnold's tongue pattern [Fig.~\ref{fig:fig5}(f)], particularly even in regions of larger $K$.


\subsection{PC Collapse in Large Inertia Limit}

At high inertia, the system exhibits a new phenomenon: the spontaneous collapse of the PC.  Specifically, for $m=12$, the system behavior can be classified into three distinct regimes based on coupling strength: no synchronization ($K < 4$), PC collapse with persistent SCs ($4 < K < 7$), and stable PC-SC coexistence ($K > 7$). For detailed analysis of these coupling regimes, see Supplementary, Sec.~\ref{sec:when $m$ is large}.

\begin{figure*}[!]
\resizebox{1.0\linewidth}{!}{\includegraphics{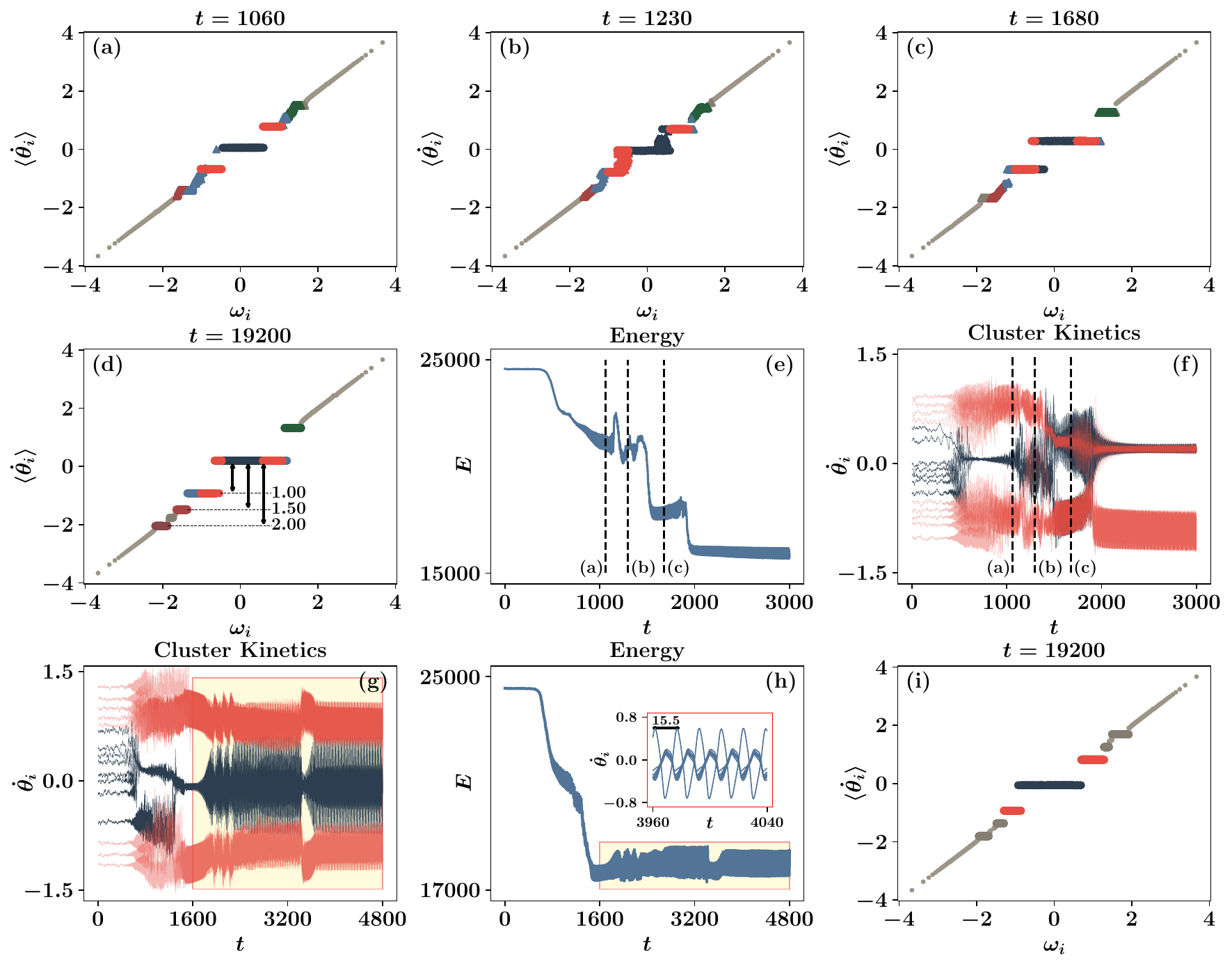}}
\centering
\caption{\textbf{Evolution of Devil's Staircase and total energy profiles.} PC collapse scenario (a)--(f): Formation of a PC (dark blue), two SCs (bright red), and HOC during the early stage (a), followed by PC collapse in the intermediate stage (b) and merging of PC fragments (dark blue) into SCs to form two enlarged SCs in the late stage (c). A Devil's Staircase pattern emerges where a reconstructed SC frequency serves as a new reference frequency, determining the rational relationships between plateaus (d). The total energy $E$ evolution shows distinct stages marked by vertical dashed lines (e). Oscillator angular velocities undergo transition phases, with PC collapse occurring at the vertical dashed line (b), where most PC fragments merge into the upper SC (f).
PC non-collapse scenario (g)--(i): Temporal dynamics showing PC fragments whose angular velocities oscillate between the values of two SCs (bright red) (g). The total energy decreases to a steady state, where the oscillation frequency of PC fragment oscillations approximates a rational multiple of the SC angular velocity (h). In the Devil's Staircase pattern of average angular velocities versus $\omega_i$, the PC appears as a plateau while its individual oscillator's actual temporal dynamics exhibit pronounced oscillations (i).}
\label{fig:fig8}	
\end{figure*}

The collapse progresses through well-defined stages [Fig.~\ref{fig:fig8}]. This spontaneous collapse results from the competition between SC's attractive forces and PC's internal cohesion. As SCs grow in strength, they draw oscillators—especially those with larger intrinsic frequencies—from the PC's periphery. These oscillators first enter an intermediate state between the PC and SCs, then merge with one of the SCs. When oscillators escape from the PC to an SC, the reduction in PC size weakens its collective stability, triggering a cascade of further departures that leads to complete PC dissolution.

\begin{figure*}[!]
\resizebox{1.0\linewidth}{!}{\includegraphics{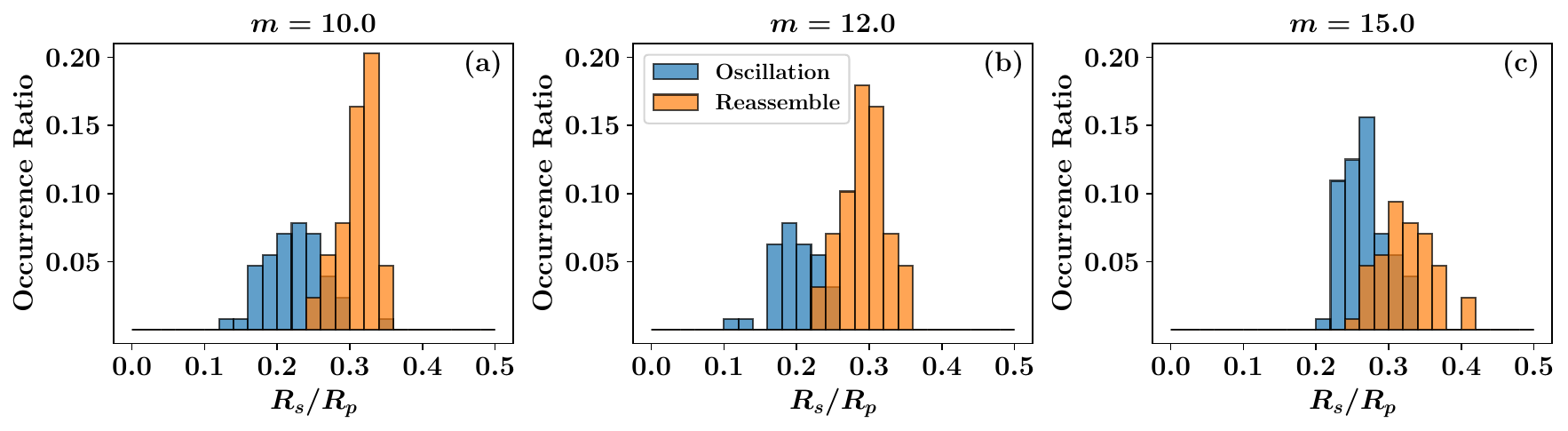}}
\caption{\textbf{Cluster size ratio determining reassembly probability.} We analyze the reassembly probabilities as a function of transient cluster size ratio ($R_{p} / R_{s}$) at coupling strength $K=6$ for (a) $m=10$, (b) $m=12$, and (c) $m=15$. Orange and blue bars indicate the probability of reassembly events and frequency oscillations, respectively. For all inertia values, higher $R_{p} / R_{s}$ values promote reassembly events, while lower $R_{p} / R_{s}$ values favor frequency oscillations. Data points represent ensemble averages over 128 independent initial configurations, with error bars indicating standard deviation. The stochastic nature of initial conditions contributes to variability in final steady states.}
\label{fig:fig9}	
\end{figure*}

We observe two distinct post-collapse behaviors depending on SC size. When SCs are sufficiently large, PC fragments completely merge into one of two SCs, yet preserve the Devil's Staircase pattern through new frequency relationships. When SCs are moderate in size, escaped oscillators execute coherent oscillations between SCs, exhibiting the frequency of these coherent oscillations at rational multiples of the SC rotational frequency. The type of post-collapse behavior is determined by the transient cluster size ratio $R_s/R_p$ [Fig.~\ref{fig:fig9}], with larger ratios leading to complete merging into SCs and smaller ratios resulting in coherent oscillations, though initial conditions introduce variability in the final state. These relationships between cluster size ratios and post-collapse dynamics provide a framework for understanding and potentially controlling the complex and diverse behaviors between primary and secondary clusters.


\section{Conclusion}
Our investigation of the second-order Kuramoto model reveals three key phenomena that transform our understanding of synchronization. These phenomena, arising from complex interactions among multiple synchronized clusters, extend beyond traditional system-wide order parameters and highlight the importance of a cluster-level analysis.

First, we show that the primary cluster (PC) interferes with secondary clusters (SCs) formation through a two-stage mechanism. Initially, the PC creates potential wells that inhibit the synchronization of candidate SC oscillators. Later, direct interactions further suppress SC formation. By tracking energy transfer from the PC to these oscillators, we observe deviations from standard Melnikov predictions. This result underscores the need to examine inter-cluster dynamics rather than relying solely on individual cluster-oscillator interactions.

Second, we observe that SCs and higher-order clusters (HOC) form near the PC’s orbital aphelion. Here, the oscillators move more slowly, resulting in prolonged inter-cluster interactions. This slower motion gives rise to Devil’s Staircase patterns, where rational frequency resonances become prominent. Consequently, our findings highlight the need for an expanded concept of clustering—one that encompasses both frequency synchronization and resonance phenomena. Furthermore, while first-order systems exhibit Bellerophon states restricted to odd-integer frequency ratios~\cite{bi2016coexistence, qiu2016synchronization, xu2018origin, qiu2019characterizing}, second-order systems exhibit a wider range of rational frequency ratios, leading to richer and more complex dynamic behavior. This complexity indicates that traditional low-dimensional reduction methods like the Ott–Antonsen ansatz~\cite{ott2008low, ott2009long, martens2009exact} cannot capture the overall dynamics of the second-order system.

Third, once SCs become sufficiently large, they can destabilize the PC and cause it to collapse. After this breakdown, PC oscillators either merge with SCs or oscillate between them. Such reorganization, driven by inter-cluster interactions, shifts the system from a metastable state to a more stable one—a phenomenon unique to inertial systems, as non-inertial systems move directly into a stable state without this intermediate reorganization.

Together, these findings advance synchronization research in two significant ways. They highlight the critical role of inter-cluster interactions, shifting the focus from traditional mean-field approaches to cluster-level analysis, and they broaden the definition of clustering to include both frequency synchronization and resonance, challenging established assumptions.

Our results have direct applications in various real-world systems exhibiting multi-cluster behavior, from shunted Josephson junction arrays to power grids and mechanical resonator networks. For example, in shunted Josephson junction arrays, multiple clusters manifest as Shapiro steps~\cite{shapiro1963josephson, lee1990calculation, kvale1991theory, lee1991quantized, rzchowski1991frequency, lee1991subharmonic, crescini2023evidence}. The link between the Devil’s Staircase and Shapiro steps reveals previously unknown cluster-level dynamics. The framework we developed helps identify and understand cluster-level phenomena in these systems, offering new insights into their operation and control.

Future research directions include; (1) developing cluster-specific phase diagrams through systematic parameter exploration, (2) extending our analytical framework to a broader range of synchronization models, and (3) investigating the impact of network topology on multi-cluster formation. These studies will help establish fundamental principles governing cluster dynamics in complex systems.

In conclusion, our study demonstrates that understanding complex synchronization phenomena requires moving beyond traditional mean-field approaches to cluster-level analysis. By revealing the crucial role of inter-cluster interactions and expanding the concept of clustering, we provide a new framework for analyzing collective behavior in dynamical systems. These insights not only advance theoretical understanding but also offer practical tools for designing and controlling synchronized systems across diverse physical and engineering applications.

During the preparation of this work the author used Claude 3.5 Sonnet and ChatGP o1 in order to language clarity and readability. After using this tool/service, the author reviewed and edited the content as needed and take full responsibility for the content of the publication.

\begin{acknowledgments}
B.K. was supported by the National Research Foundation of Korea by Grant No. RS-2023-00279802 and the KENTECH Research Grant No. KRG-2021-01-007.
S.B. acknowledges support from the project n.PGR01177 of the Italian Ministry of Foreign Affairs and International Cooperation.
\end{acknowledgments}



\bibliographystyle{apsrev4-2}
\bibliography{MCD_CSF_Ref}

\clearpage
\newpage

\appendix

\onecolumngrid


\makeatletter
\renewcommand{\thesection}{S\arabic{section}}
\renewcommand{\theequation}{S\arabic{equation}}
\renewcommand{\thefigure}{S\arabic{figure}}
\renewcommand{\thetable}{S\arabic{table}}
\renewcommand{\bibnumfmt}[1]{[S#1]}

\setcounter{section}{0}
\setcounter{equation}{0}
\setcounter{figure}{0}
\setcounter{table}{0}
\setcounter{page}{1}
\makeatother

\begin{center}
\textbf{\large Supplementary Material for \\ Cluster-Mediated Synchronization Dynamics in \\ Globally Coupled Oscillators with Inertia}
\end{center}


\section{The Melnikov method: PC boundary}
\label{sec:The Melnikov method: PC boundary}

Here, we explain the derivation of the term $(4 \gamma/{\pi})\sqrt{K R_{a}/m}$ that appears in Eqs.~\eqref{eq:omega_p_range} and~\eqref{eq:omega_s_range} of the main text, where $a=p$ or $s$.
Consider an oscillator $i$ rotating with a finite angular velocity $\dot \theta_i \ne 0$. The oscillator's rotational motion can be maintained only if the total non-conservative energy ($E_{\rm nc}$) remains non-negative:
\begin{align}
E_{\rm nc} \equiv W_{\omega} + W_{\gamma} \geq 0,
\end{align}
where $W_\omega$ represents the total energy supplied by the oscillator's intrinsic frequencies, and $W_\gamma$ denotes the total energy loss due to dissipation.
The challenge in applying this energy criterion lies in calculating $E_{\rm nc}$ for nonlinear dynamics, where precise trajectory calculations become impossible beyond a certain time limit.
This limitation makes it impossible to calculate the path-dependent quantities $W_\omega$ and $W_\gamma$, which are essential components of $E_{\rm nc}$.
The Melnikov method (MM) addresses this challenge by analyzing the nonlinear dynamics in the limit $E_{\rm nc}\to 0$, where the trajectory remains nearly unchanged from the unperturbed path, making the analysis tractable.
Let ($\theta(t)$, $\dot \theta(t)$) represent the angle and angular velocity configurations when $E_{\rm nc}=0$. When $\omega$ and $\gamma$ are activated, the rotational motion trajectory slightly shifts to ($\theta_{\omega, \gamma}(t)$, $\dot \theta_{\omega, \gamma}(t)$), where $\theta_{\omega, \gamma}\approx \theta$ and $\dot \theta_{\omega, \gamma}\approx \dot \theta$. In this unperturbed state, the angular velocity follows the relationship:
\begin{align}
\dot{\theta} = \sqrt{\dfrac{2KR_a}{m}(1 - \cos{\theta})},
\label{eq:trajectory_0}
\end{align}
which describes the conservative motion of the oscillator. During one revolution, the energy change due to the perturbation is given by:
\begin{align}
E_{\rm nc}
& = \int_{0}^{2\pi} d \theta \omega - \int_{0}^{2\pi} d \theta \gamma \dot{\theta} \cr
& = 2\pi \omega - 4\gamma \int_{-\infty}^{\infty} dt \dfrac{K R_a}{m} \sech^2(\sqrt{\dfrac{K R_a}{m}}t) \cr
& = 2\pi \omega - 8 \gamma \sqrt{\dfrac{K R_a}{m}},
\end{align}
where the second line is obtained by transforming to the time domain using Eq.~\eqref{eq:trajectory_0}.
The oscillator stops when $E_{\rm nc} < 0$, or equivalently, when $\omega \le (4 \gamma/{\pi})\sqrt{K R_{a}/m}$. This criterion extends to multi-oscillator systems: for oscillators with intrinsic frequencies ${ \omega_i }$ in cluster $a$ with mean angular velocity $\omega_{\bar a}$, we require:
\begin{align}
\omega - \omega_{\bar a} \le \dfrac{4 \gamma}{\pi} \sqrt{\dfrac{K R_a}{m}}.
\end{align}
While this MM-derived upper bound captures the essential physics, it shows a slight deviation from numerical results. To improve accuracy, we follow Ref.~\cite{gao2018self} and introduce a correction term $\delta \omega_a$ that accounts for effects beyond the MM approximation.


\clearpage
\newpage


\section{{\bfadhoc} potential of PC}
\label{sec:adhoc potential of PC}
\begin{figure}[h]
\resizebox{\columnwidth}{!}{\includegraphics{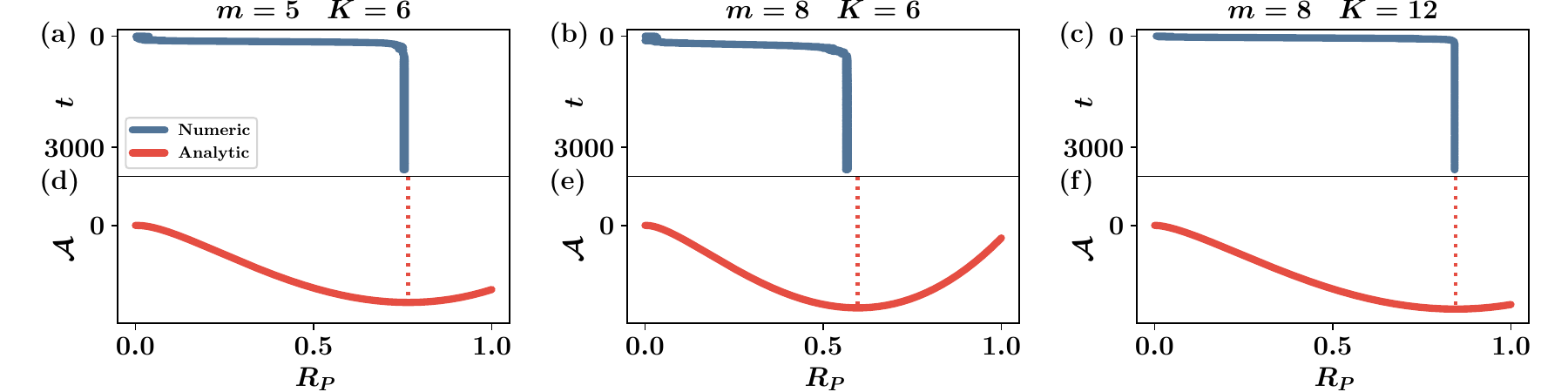}}
\caption{\textbf{Evolution of PC size and validation of modified MM prediction.} Analogous to [Fig.~\ref{fig:fig1}] in the main text, we present results for different parameter sets: $(m,K)=$ (5,6) (a,d), (8,6) (b,e), and (8,12) (c,f). (a) Temporal evolution of $R_P$, the fraction of oscillators in PC. As time progresses (top to bottom), $R_P$ converges to its steady-state value $R_P(\infty)$. (d) The \adhoc potential $\mathcal{A}(R_P)$ versus $R_P$, calculated using the analytic formulae from Eqs.~\eqref{eq:omega_p_range} and~\eqref{eq:adhoc} of the main text (bright red curve). The minimum of $\mathcal{A}(R_P)$ occurs at $R_P^*$ (bright red), which closely matches $R_P(\infty)$ (dotted line). This agreement validates the MM prediction with correction term $\delta \omega$ for determining the $\omega_p$ range boundary. Panels (b,e), and (c,f) show corresponding results for the other parameter sets.
}
\label{fig:fig_S_1}	
\end{figure}

[Fig.~\ref{fig:fig_S_1}] shows the \adhoc potential of PC for different sets ($m,K$). The $\omega$ boundary of the PC cluster is determined using the MM with correction $\delta \omega$, which works well to match $R_P(\infty)$ with $R_P^*$ (see the main text for definitions).

\clearpage
\newpage


\section{SC forms after PC growth}
\label{sec:SC forms after PC growth}

\begin{figure}[h]
\resizebox{1.0\columnwidth}{!}{\includegraphics{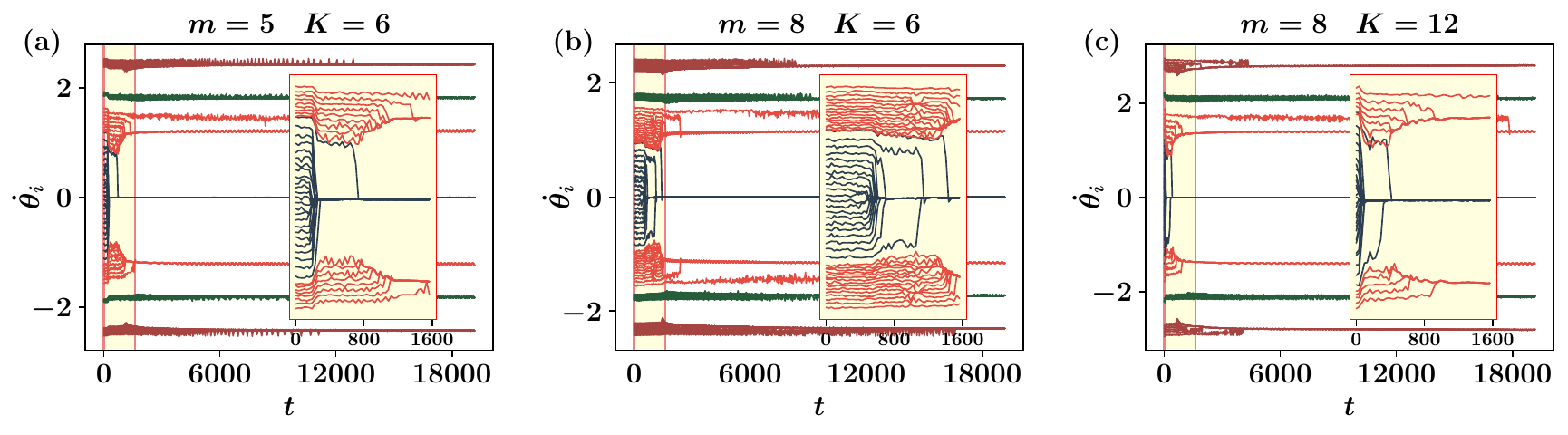}}
\caption{\textbf{Evolution of angular velocities $\dot \theta_i(t)$.} For different parameter sets $(m,K)=$ (5,6) (a), (8,6) (b), and (8,12) (c). Oscillators are color-coded by their cluster type: PC (dark blue), SC (bright red), TC (dark green), and QC (deep red). Inset: Extended time view ($t\in (0,1200)$) showing the early formation of PC relative to other clusters.
}
\label{fig:fig_S_2}	
\end{figure}

[Fig.~\ref{fig:fig_S_2}] illustrates how the PC (whose oscillators are marked in blue) is formed at an early stage, followed by the formation of other clusters. 

\subsection{SC growth is suppressed}
\label{subsec:SC growth is suppressed}

We analyze the system dynamics when the PC grows but before the SCs form. During this phase, any oscillator $i$, whether in $C_p$ or $C_s$, can rotate periodically around the PC's center of mass.
Following Ref.\cite{gao2018self}, we can express the angular velocity $\dot \theta_i$ in terms of $\sin\theta_{i, \bar{p}}$ and $\cos \theta_{i,\bar{p}}$:
\begin{align}
\dot{\theta}_{i} \simeq (\omega_{i}/\gamma) + C_{i,1} \cos \theta_{i,\barp} + S_{i,1} \sin \theta_{i,\barp} + \text{h.o.},
\label{eq:C1}
\end{align}
where higher-order terms are negligible. By substituting this expression into Eq.~\eqref{eq:E-M_p} of the main text and collecting the coefficients of $\sin\theta_{i,\barp}$ and $\cos\theta_{i,\barp}$ terms, we obtain:
\begin{align}
m (\omega_{i}/\gamma) S_{i,1} + \gamma C_{i,1} & \simeq  0 & \quad \textrm{for } & \quad  \cos \theta_{i,\barp}, \cr
m (\omega_{i}/\gamma) C_{i,1} + \gamma S_{i,1} & = K R_{P} & \quad \textrm{for } & \quad  \sin \theta_{i,\barp}.
\end{align}
Solving these equations yields the coefficients $C_{i,1}$ and $S_{i,1}$:
\begin{align}
C_{i,1} \simeq \dfrac{m \omega_{i}}{\gamma} \dfrac{K R_{P}}{\gamma^2 + (m \omega_{i} / \gamma)^2}, S_{i,1} \simeq - \gamma \dfrac{K R_{P}}{\gamma^2 + (m \omega_{i} / \gamma)^2}.
\label{eq:Coeff_C_S}
\end{align}
Substituting these coefficients back into Eq.~\eqref{eq:C1} and taking the time average, we find:
\begin{align}
\langle \dot{\theta}_{i} \rangle{t} \simeq (\omega_{i}/\gamma) + C_{i,1} \langle \cos\theta_{i,\barp} \rangle_t + S_{i,1} \langle \sin\theta_{i,\barp} \rangle_t \simeq (\omega_{i}/\gamma) - \left(\dfrac{m^2 \omega_{i}}{2\gamma} + \dfrac{\gamma^3}{2\omega_{i}}\right) \left( \dfrac{K R_{P}}{\gamma^2 + (m \omega_{i} / \gamma)^2} \right)^2,
\end{align}
where the time averages are given by:
\begin{align}
\langle \sin (\theta_{i,\barp}) \rangle_t
& = \dfrac{1}{T} \int_{0}^{T} \sin\theta_{i,\barp}(t) dt = \dfrac{1}{2 \pi} \int_{0}^{2 \pi} \dfrac{\sin\theta_{i,\barp}}{\dot{\theta}_{i,\barp}} d\theta_{i,\barp} \Bigg{/} \dfrac{1}{2 \pi} \int_{0}^{2 \pi} \dfrac{1} {\dot{\theta}_{i,\barp}} d\theta_{i,\barp} \cr
\langle \cos(\theta_{i,\barp}) \rangle_t
& = \dfrac{1}{T} \int_{0}^{T} \cos\theta_{i,\barp}(t) dt = \dfrac{1}{2 \pi} \int_{0}^{2 \pi} \dfrac{\cos\theta_{i,\barp}}{\dot{\theta}_{i,\barp}} d\theta_{i,\barp} \Bigg{/} \dfrac{1}{2 \pi} \int_{0}^{2 \pi} \dfrac{1} {\dot{\theta}_{i,\barp}} d\theta_{i,\barp}
\end{align}
For $m \ge 5$ with $\gamma = 1$, the inertia effect dominates over dissipation, allowing us to approximate the time-averaged angular velocity as:
\begin{align}
\langle \dot{\theta}_{i} \rangle_{t} \simeq \omega_{i}- \dfrac{m^2 \omega_{i}}{2} \left( \dfrac{K R_{P}}{1+ (m \omega_{i})^2} \right)^2 \simeq \omega_{i} - \dfrac{(K R_{P})^2}{2 m^2 \omega_{i}^3} \equiv \omega_{i} - \delta \dot{\theta}_{i}.
\end{align}
Thus, as time progresses, the average angular velocity of a rotating oscillator decreases from its initial value $\omega_i$ by $\delta \dot{\theta}_i$, which scales linearly with $R_{P}^{2}$ and inversely with $\omega_{i}^{3}$.
When the SC begins to form, oscillators $j\in C_s$ have an average angular velocity in the reference frame of ${\dot{\theta}}_{\barp}=0$ given by:
\begin{align}
\langle \dot{\theta}_{j} \rangle \simeq \dfrac{\omega_{j}}{\gamma} - \dfrac{(K R_{P})^2}{2 m^2 (\omega_{j}/\gamma)^3}.
\end{align}
The SC's center of mass rotates with angular velocity:
\begin{align}
\langle \dot{\theta}_{\bar{s}} \rangle \simeq \dfrac{\omega_{\bar{s}}}{\gamma} - \dfrac{(K R_{P})^2}{2 m^2 (\omega_{\bar{s}}/\gamma)^3}.
\end{align}
Consequently, the relative average velocity of oscillator $j$ with respect to $\dot{\theta}{\bars}$ becomes:
\begin{align}
\langle \dot{\theta}_{j, \bar{s}} \rangle_{t}
& \simeq \left(\dfrac{\omega_{j}}{\gamma} - \dfrac{\omega_{\bar{s}}}{\gamma} \right) - \dfrac{(K R_{P})^2}{2m^2} \left( \dfrac{1}{(\omega_{j}/\gamma )^3} - \dfrac{1}{(\omega_{\bar{s}}/\gamma )^3} \right) \cr
& \equiv \left(\dfrac{\omega_{j}}{\gamma} - \dfrac{\omega_{\bar{s}}}{\gamma} \right) + \delta \dot{\theta}_{j,\bars}.
\end{align}
The correction term $\delta \dot{\theta}_{j,\bars}$ is negative when $\omega_j < \omega_{\bars}$ and positive when $\omega_j > \omega_{\bars}$. As $R_P$ increases, this leads to increasing deviation $|\delta \dot{\theta}_{j, \bar{s}}|$ from $\omega_{\bars}$, making it progressively more difficult for oscillator $j$ to merge with the SC.


\subsection{How to measure cluster size in transient regime}
\label{subsec:How to measure cluster size in transient regime}

\begin{figure}[h]
\resizebox{\columnwidth}{!}{\includegraphics{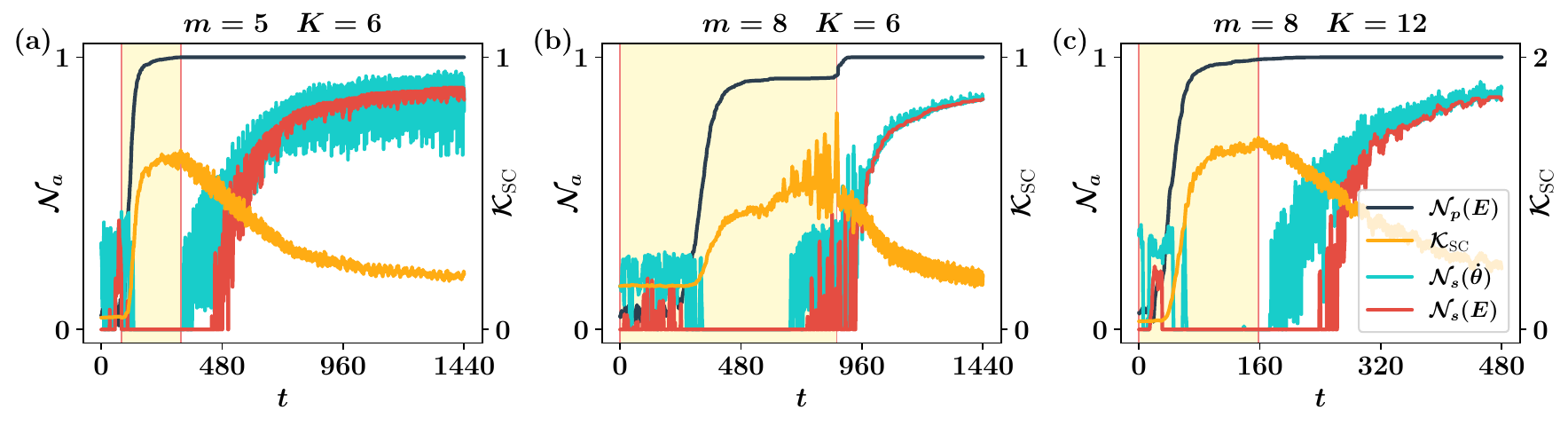}}
\caption{\textbf{Evolution of PC and SC normalized sizes.} For different parameter sets $(m,K)=$ (5,6) (a), (8,6) (b), and (8,12) (c). We plot the number of oscillators in PC and SCs, normalized by their final values: $\mathcal{N}_{p}(E,t)=N_p(E,t)/N_p(E,\infty)$ for PC using the energy criterion (dark blue), $\mathcal{N}_{s}(E,t)=N_s(E,t)/N_s(E,\infty)$ for SC using the energy criterion (bright red), and $\mathcal{N}_{s}(\dot{\theta},t)=N_s(\dot{\theta},t)/N_s(\dot{\theta},\infty)$ for SC using the angular velocity criterion (cyan). The internal kinetic energy of SC, $\mathcal{K}_{SC}$, is also shown (yellow). The results reveal that $\mathcal{K}_{SC}$ increases during PC growth and reaches its maximum when the PC growth rate becomes zero. At this point, the SC begins to form, showing explosive growth characteristic of explosive percolation~\cite{achlioptas2009explosive, d2015anomalous, cho2016hybrid, choi2017critical}.
}
\label{fig:fig_S_3}	
\end{figure}

To understand the formation dynamics of the SC, we analyze the trajectory ${\theta_j(t), \dot \theta_j(t)}$ of each potential SC member (oscillator $j\in C_s$) over time. The process occurs in distinct phases. Initially, these oscillators experience a strong attraction toward the PC's center, with the attracting torque proportional to the PC's growth rate $dR_{P}/dt$. This attraction ceases when the PC completes its growth ($dR_{P}/dt = 0$), marking the onset of SC formation.
To quantify these dynamics, we introduce the internal kinetic energy of the SC:
\begin{equation}
{\mathcal K}_{\text{SC}} = \sum_{j \in C_s} \dfrac{1}{2} m \left( \dot{\theta}_{j} - \dot{\theta}_{\bar s} \right)^2.
\label{eq:kinetic-energy-sc}
\end{equation}
The evolution of ${\mathcal K}_{\text{SC}}$, shown in [Fig.~\ref{fig:fig_S_3}], reveals that it reaches its maximum value when PC growth completes, then decreases as oscillators begin clustering around the SC's center, leading to rapid SC growth.
We track the SC's growth by measuring the number of merged oscillators ($N_S$) using two distinct criteria:
\begin{itemize}
\item \textit{Total Energy Criterion:}
We calculate each oscillator's energy relative to the SC:
\begin{equation}
{\mathcal E}_{j,s}= \dfrac{1}{2} m (\dot{\theta}_j-\dot \theta_{\bars})^2 - K R_{S} \cos(\theta_j-\theta_{\bars}).
\label{eq:total-energy}
\end{equation}
An oscillator is considered part of the SC when $\mathcal{E}_{j,s} < 0$. Initially, all oscillators have $\mathcal{E}_{j,s} > 0$, but as time progresses, $\mathcal{E}_{j,s}$ decreases until reaching a saturation time $t_s$ when the SC is complete.
\item \textit{Frequency Entrainment Criterion:}
Alternatively, we identify SC members based on their angular velocity:
\begin{equation}
\left|  \langle\dot{\theta}_j\rangle- \langle \dot{\theta}_{\bars} \rangle\right| < \epsilon,
\label{eq:kinetic_epsilon}
\end{equation}
where $\epsilon = 10^{-2}$ serves as a threshold. Oscillators satisfying this condition are counted as SC members.
\end{itemize}
The temporal evolution of $N_S$ measured by both criteria is presented in [Fig.~\ref{fig:fig_S_3}].

\begin{figure}[h!]
\resizebox{\columnwidth}{!}{\includegraphics{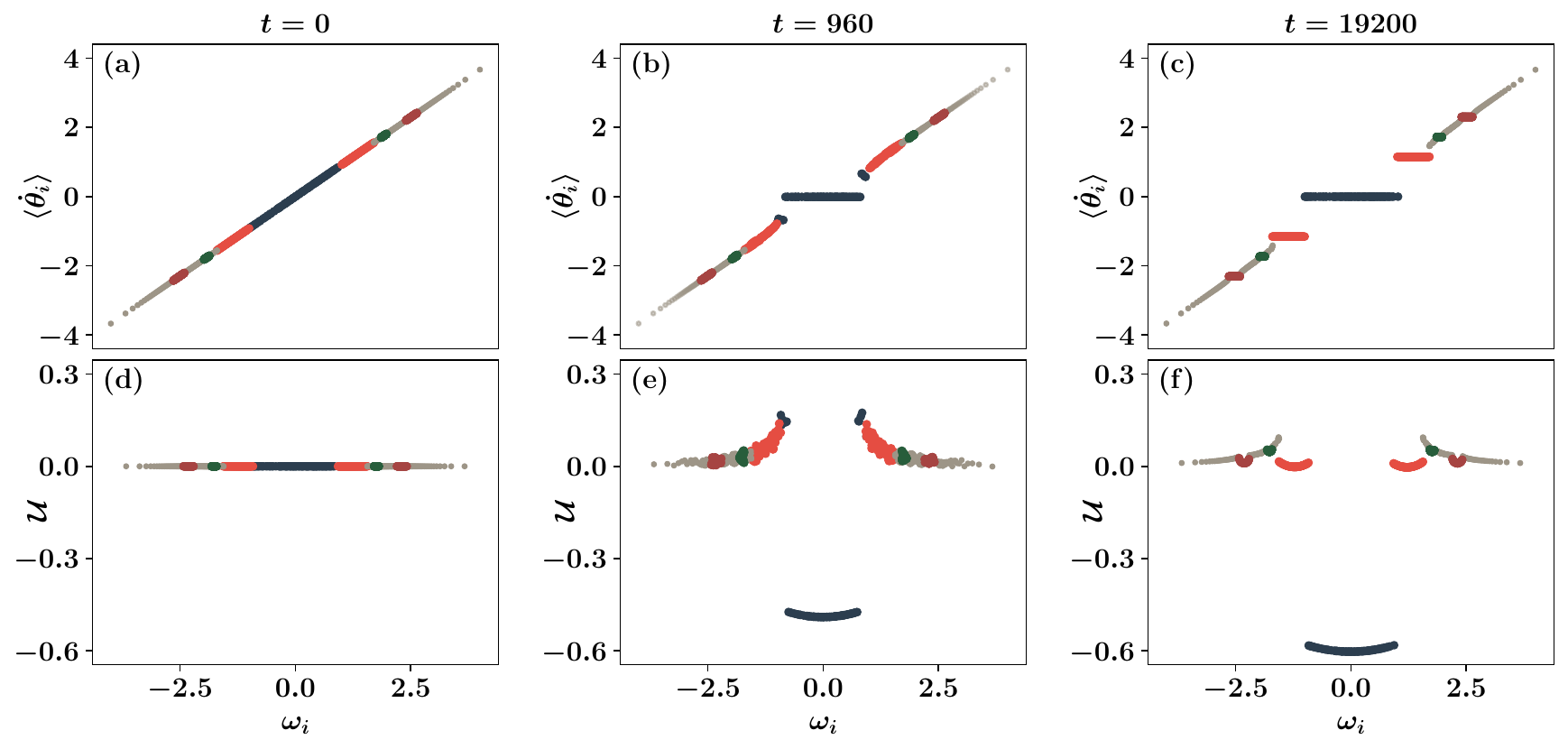}}
\caption{\textbf{Evolution of average angular velocities and potential energy landscapes.}
(a) Initial state ($t=0$) showing uniform potential energy distribution $\mathcal{U(\omega)}$ across all frequencies.
(b) Formation of PC potential well at $\omega=0$ during intermediate stage ($t=960$).
(c) Fully developed potential landscape at a later stage ($t=19200$) exhibiting hierarchical structure with a global minimum for PC (dark blue) and local minima for SCs (bright red), TCs (dark green), and QCs (deep red).
(d)--(f) Average angular velocity $\langle \dot \theta_i \rangle$ versus intrinsic frequency $\omega_i$ corresponding to the times shown in (a)--(c). The averages were computed over time windows centered at each specified time point.}
\label{fig:fig_S_4}	
\end{figure}

Next, we analyzed the temporal evolution of the potential energy profile and the associated Devil's Staircase pattern. The total potential energy of the system can be expressed as:
\begin{align}
\mathcal{U}(\theta_i(t))= -
\sum_a KR_{a}\cos{(\theta_i(t) - \theta_{\bar a})},
\label{eq:potential_total}
\end{align}
where the index $a$ encompasses various clustering states (primary, secondary, tertiary, etc.). By calculating $\theta_i(t)$ for all oscillators $i$, we tracked the evolution of the potential energy profile as a function of natural frequency $\omega$ at different time points [Fig.~\ref{fig:fig_S_4}]. Our analysis revealed a sequential formation of potential wells, beginning with PC, followed by the emergence of additional wells corresponding to SCs, TCs, and QCs. This hierarchical formation process manifests as distinct plateaus in the frequency domain.

\clearpage
\newpage

\clearpage
\newpage


\section{SC boundary}
\label{sec:SC boundary}

We return to the coarse-grained Kuramoto equation~\eqref{eq:E-M_s} for an oscillator $j$ in SC:
\begin{align}
m \ddot{\theta}_{j,s} + \gamma \dot{\theta}_{j,s} = (\omega_j - \omega_{\bar{s}}) + \tau_{s,j} + \tau_{p,j} - \tau_{p,s}.
\label{eq:single_rotor_secondary_cluster}
\end{align}
Here, $(\omega_j- \omega_{\bar{s}})$ and $\tau_{s,j}$ represent the primary torques determining the oscillator's steady state in the absence of the PC. The additional terms $\tau_{p,j} = KR_{P}\sin \theta_{\barp, j}$ and $\tau_{p,s} = KR_{P} \sin \theta_{\barp, \bars}$ denote the torques exerted by the PC on oscillator $j$ and the SC's center of mass, respectively. Given that $R_{P} \gg R_{S}$, these PC-induced torques significantly influence SC formation.
The energy change ($\Delta E_{p,j}$) of oscillator $j$ due to these torques can be expressed as:
\begin{align}
\Delta E_{p,j}
& = \int_{0}^{2\pi} d \theta_{j, \bars} \left(\tau_{p,j} - \tau_{p,s}\right) \cr
& = \int_{0}^{2\pi} d \theta_{j, \bars} K R_{P} \left[\sin(\theta_{\barp, j}) - \sin(\theta_{\barp, \bars})\right].
\label{eq:work_primary_cluster}
\end{align}
Direct calculation of Eq.~\eqref{eq:work_primary_cluster} is challenging due to the unknown dependence of $\theta_{\barp,j}$ and $\theta_{\barp, \bars}$ on $\theta_{j, \bars}$. However, we can simplify this expression by considering the high angular velocities of oscillators $j$ and $s_{\cm}$. During one orbit of $j$ around the SC, multiple encounters with the PC occur, implying that $\theta_{\barp,j}$ and $\theta_{\barp,\bars}$ span the full range $[0, 2\pi]$ for any fixed $\theta_{j,\bars}$. This allows us to use time-averaged values:
\begin{align}
\Delta E_{p,j}
& \simeq \int_{0}^{2\pi} d \theta_{j, \bars} K R_{P} \left[\langle \sin(\theta_{\barp, j}) \rangle - \langle \sin(\theta_{\barp, \bars}) \rangle \right],
\label{eq:T-AV_sin}
\end{align}
Using methods from Sec.~\ref{sec:SC forms after PC growth} and Ref.~\cite{gao2018self}, we obtain:
\begin{align}
\langle\sin(\theta_{\barp,j})\rangle \simeq -\dfrac{\gamma K R_{P}}{2m^2(\omega_{j}/\gamma)^3}, \langle\sin(\theta_{\barp,\bars})\rangle \simeq -\dfrac{\gamma K R_{P}}{2m^2(\omega_{\bars}/\gamma)^3}.
\end{align}
This leads to the energy change in the $s_{\cm}$-reference frame:
\begin{align}
\Delta E_{p,j}
& = \int_{0}^{2\pi} d \theta_{j,\bars} K R_{P} \left[\langle \sin(\theta_{\barp,j}) \rangle - \langle \sin(\theta_{\barp, \bars}) \rangle \right], \cr
& \simeq \int_{0}^{2\pi} d \theta_{j, \bars} \gamma \dfrac{(K R_{P})^2}{2 m^2} \left[- \dfrac{1}{(\omega_{j}/\gamma)^3} + \dfrac{1}{(\omega_{\bars}/\gamma)^3} \right].
\label{eq:energy_change_PC_SM}
\end{align}
Following a similar approach to that used for $\delta \omega_p$, we derive corrections to the $\omega_s$ boundary:
\begin{align}
\Delta \omega_{s, \max}
& = \dfrac{\beta}{2} \gamma \left( \dfrac{K R_{P}}{m} \right)^2 \left[\dfrac{1}{(\omega_{\bars}/\gamma)^3} - \dfrac{1}{(\omega_{j}/\gamma)^3}\right] \cr
& \simeq \dfrac{\beta}{2} \gamma \left( \dfrac{K R_{P}}{m} \right)^2 \left[\dfrac{1}{\omega_{\bars}^3} - \dfrac{1}{\left(\omega_{\bars} + 4\gamma/\pi \sqrt{K R_{S} / m}\right)^3}\right] \cr
\Delta \omega_{s, \min}
& = \dfrac{\beta}{2} \gamma \left( \dfrac{K R_{P}}{m} \right)^2 \left[\dfrac{1}{(\omega_{j}/\gamma)^3} - \dfrac{1}{(\omega_{\bars}/\gamma)^3}\right] \cr
& \simeq \dfrac{\beta}{2} \gamma \left( \dfrac{K R_{P}}{m} \right)^2 \left[\dfrac{1}{\left(\omega_{\bars} - 4\gamma/\pi \sqrt{K R_{S} / m}\right)^3} - \dfrac{1}{\omega_{\bars}^3}\right],
\label{eq:omega_secondary_change}
\end{align}
where $\beta$ is a constant to be determined, analogous to $\alpha$ in the PC correction term $\delta \omega_p$. Here, $\Delta \omega_{s, \max}$ and $\Delta \omega_{s, \min}$ represent the PC-induced corrections to the SC frequency boundaries $\omega_{s, \max}$ and $\omega_{s, \min}$.


\clearpage
\newpage


\section{\bfadhoc potential for SC}
\label{sec:adhoc potential for SC}

\begin{figure}[h]
\resizebox{\columnwidth}{!}{\includegraphics{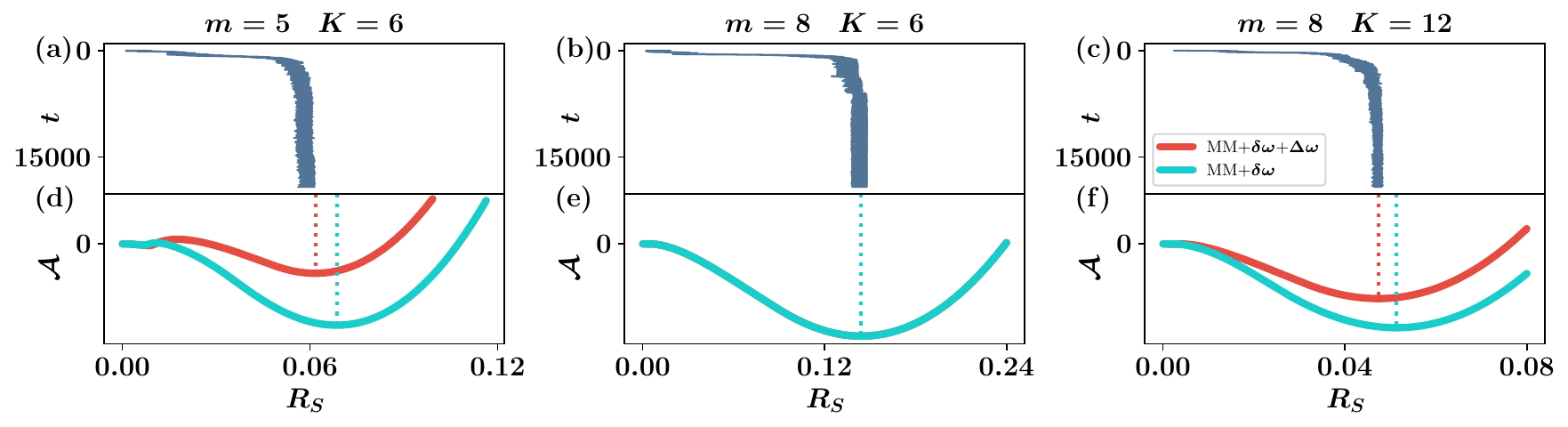}}
\caption{\textbf{Evolution of SC size and necessity of extended MM corrections.}
(a)--(c) Evolution of the SC order parameter $R_S(t)$ for different parameter sets ($m, K$). The curves show convergence to steady-state values $R_S(\infty)$ as time increases (downward direction on $y$-axis).
(d)--(f) \adhoc potential $\mathcal{A}(R_S)$ versus $R_S$ calculated using two approaches: our proposed method including both corrections $\delta \omega+\Delta \omega$ (bright red curves) and the conventional approach with only $\delta \omega$ correction (cyan curves). Note the complete overlap of both curves in panel (e). Each approach yields a characteristic minimum at $R_S^*$.
The cyan $R_S^*$ values deviate from the numerically obtained $R_S(\infty)$ shown in (a)--(c), indicating that the conventional approach alone cannot accurately predict the $\omega_s$ boundaries. Our analysis demonstrates the necessity of including the additional correction term $\Delta \omega$, with its magnitude determined by the coefficient $\beta$. This coefficient depends on both system parameters ($m$, $K$) and initial conditions. The fitted values are $\beta = 0.844$, $0.0$, and $0.673$ for panels (a,b,c), respectively, closely matching the statistically estimated values: 
$\beta^* = 0.833 \pm 0.022$ for $(m,K) = (5,6)$, $\beta^* = 0.036 \pm 0.108$ for $(m,K) = (8,6)$, and $\beta^* = 0.673 \pm 0.052$ for $(m,K) = (8,12)$. We observe that $\beta$ increases with decreasing $m$ or increasing $K$, while remaining below unity ($\beta^* < 1$), indicating a moderated $R_p$ effect compared to the case $\beta=1$. The precise functional dependence of $\beta^*$ on ($m$, $K$) remains an open question for future investigation.}
\label{fig:fig_S_5}	
\end{figure}

[Fig.~\ref{fig:fig_S_5}] illustrates the dynamics and stability of the SC for three different parameter sets: $(m,K)$ = (5,6), (8,6), and (8,12). The upper panels (a)--(c) show the temporal evolution of the SC order parameter $R_S$, while the lower panels (d)--(f) display the corresponding \adhoc potential as a function of $R_S$. The theoretical boundaries for SC formation, derived from Eq.~\eqref{eq:omega_s_range} with corrections $\delta \omega$ and $\Delta \omega$, are shown in orange. Notably, the equilibrium values $R_S(\infty)$ observed in the temporal evolution (a)--(c) align precisely with the potential minima $R_S^*$ (d)--(f), validating our previously proposed approach (Supplementary~\ref{sec:SC boundary}).

\clearpage
\newpage


\section{The case of a small inertial: Absence of Secondary Clusters}
\label{sec:when $m$ is small}

\begin{figure}[h]
\resizebox{\columnwidth}{!}{\includegraphics{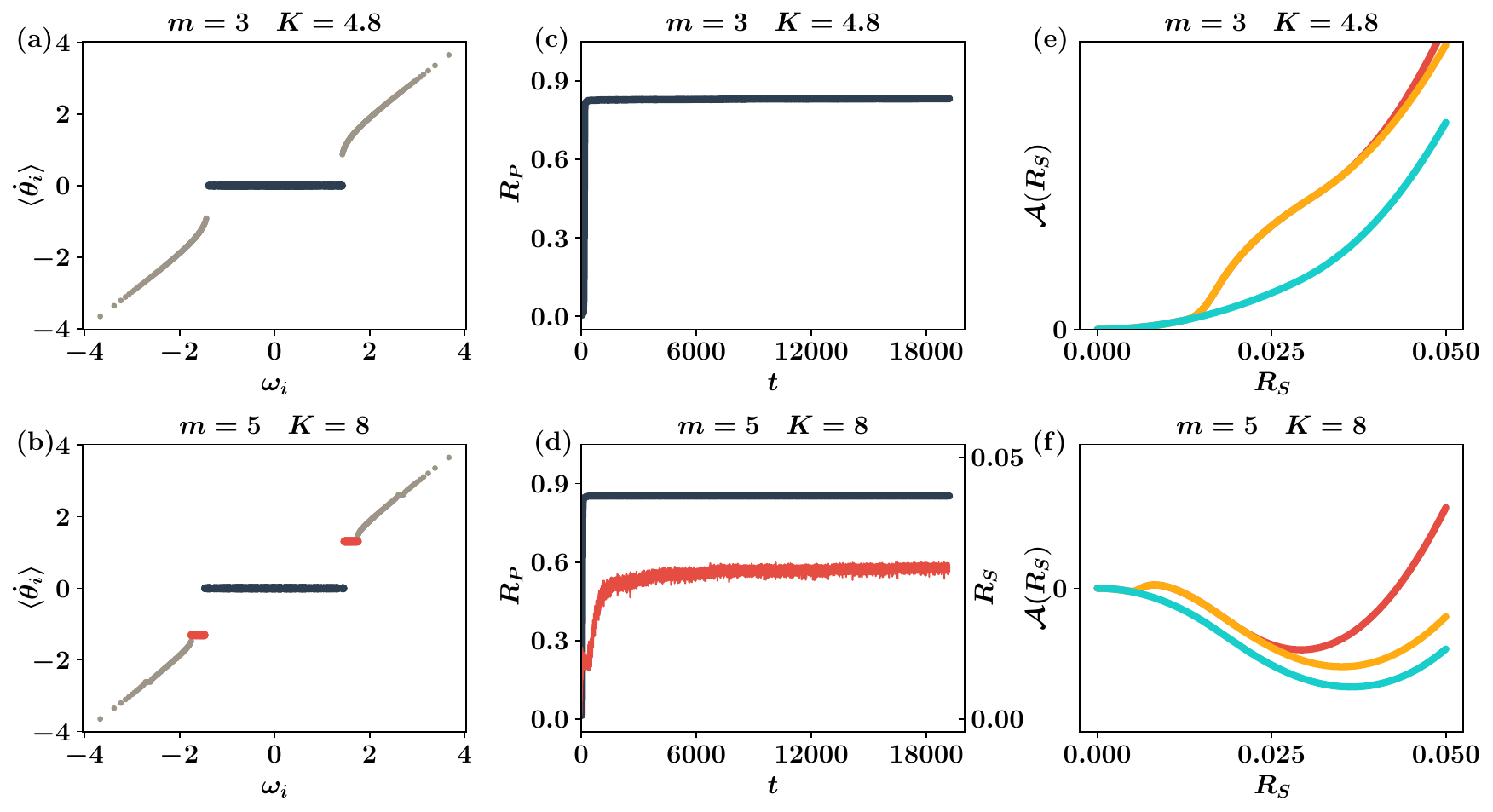}}
\caption{\textbf{Mass-dependent emergence of \bfadhoc potential landscapes in synchronized clusters.} (a,b) Time-averaged angular velocity $\langle \dot \theta \rangle_t$ versus natural frequency $\omega$, (c,d) temporal evolution of partial synchronization order parameter $R_P$, and (e,f) \adhoc potential landscapes as functions of $R_S$. Results correspond to parameter sets $(m,K)=(3, 4.8)$ (left panels) and $(5,8)$ (right panels). A local potential well emerges only for the larger $m$ value.}
\label{fig:fig_S_8}
\end{figure}

We demonstrate that the $\omega_s$ window~\eqref{eq:omega_s_range} presented in the main text represents a necessary but insufficient condition for SC formation. To illustrate this, we analyzed two distinct cases: $(m,K)=(3,4.8)$ and $(5,8)$. [Fig.~\ref{fig:fig_S_8}] reveals that SC formation occurs in the latter case $(5,8)$ but not in the former $(3,4.8)$. This conclusion is supported by our analysis of three key metrics: $\langle {\dot \theta_i} \rangle$ versus $\omega_i$, the temporal evolution of $R_a$ where $a\in (p,s)$, and $\mathcal{A}(R_S)$ versus $R_s$. Our findings indicate the existence of a characteristic threshold $m_*$, below which SC formation is impossible even when $\omega_s$ falls within the boundary defined by Eq.~\eqref{eq:omega_s_range} in the main text.

\clearpage
\newpage


\section{Devil's Staircase Pattern}
\label{sec:Devil's Staircase Pattern}

Our numerical analysis reveals an intriguing Devil’s Staircase pattern in the multi-cluster dynamics. As shown in Table~\ref{ta:frequency_ratio}, the frequency ratios measured for various mass and coupling-strength parameters $(m,K)$ exhibit universal values. Notably, these ratios consistently converge to rational multiples—namely 1.5 and 2.0—across all 128 independent realizations, each starting from a distinct initial configuration. This demonstrates that the system’s dynamics display robust frequency resonance behavior independent of initial conditions.

\begin{table}[h]
\centering
\begin{normalsize}
\setlength{\tabcolsep}{5pt}
{\renewcommand{\arraystretch}{1.8}
\begin{tabular}{@{\extracolsep{\fill}}cccc}
\hline
\hline
$(m,K)$&\multicolumn{1}{c}{(5,6)}&\multicolumn{1}{c}{(8,6)}&\multicolumn{1}{c}{(8,12)} \cr
\hline
{TC}&$1.500146$&$1.499978$&$1.499904$ \cr
{QC}&$2.000173$&$1.999960$&$1.999863$ \cr
\hline
\end{tabular}}
\end{normalsize}
\caption{\textbf{Universal frequency ratios in multi-cluster dynamics.} The ratios $\langle \dot{\theta}_{\bar t} \rangle / \langle \dot{\theta}_{\bar s} \rangle$ and $\langle \dot{\theta}_{\bar q} \rangle / \langle \dot{\theta}_{\bar s} \rangle$ were measured for various parameter sets $(m,K)$, showing values closely matching 1.5 and 2, respectively. Results represent ensemble averages over 128 distinct initial configurations.
\label{ta:frequency_ratio}
}
\end{table}

\clearpage
\newpage


\section{SC effects on HOCs}
\label{sec:SC effect on the HOC}

\begin{figure}[h]
\resizebox{1.0\columnwidth}{!}{\includegraphics{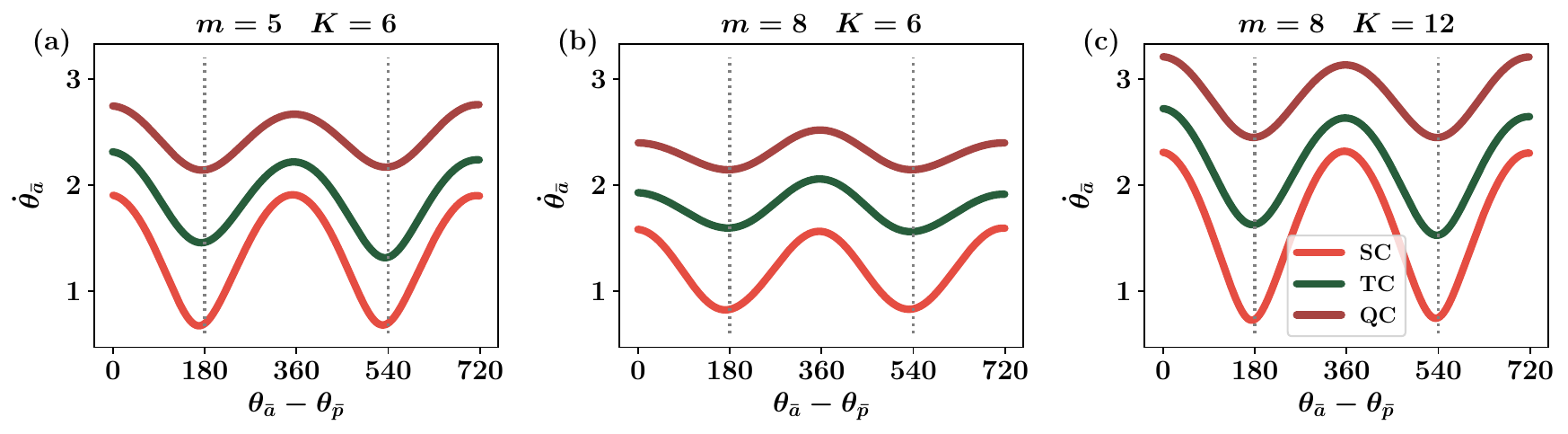}}
\caption{\textbf{Angular velocity variations of SCs and HOCs relative to the PC.} For parameter sets $(m,K)=$ (5,6) (a), (8,6) (b), and (8,12) (c). Each panel shows the angular velocity $\dot \theta_{\bar a}$ of cluster $a$ (where $a \in {S,T,Q}$ denotes SCs, TCs, QCs, respectively) versus the phase difference $\theta_{\bar a}-\theta_{\bar p}$ between the center of mass of cluster $a$ and the PC. All clusters exhibit minimum angular velocity at the aphelion position relative to the PC.}
\label{fig:fig_S_6}	
\end{figure}

To understand the formation mechanism of HOCs, we analyze a specific case where TC oscillators complete three rotations around the PC for every two rotations of the SC. In this scenario, TC oscillators interact with the SC once every three revolutions. Due to the relatively small size of the TC compared to the PC and SC, TC oscillators experience weaker self-attraction within their cluster. Instead, they are predominantly influenced by the SC near the aphelion of the PC, where they move slowly due to their low kinetic energy.
The potential energy between oscillator $j\in C_T$ and the PC's center of mass is given by:
\begin{align}
\mathcal{U}(\theta_j)= -KR_{P}\cos{(\theta_j - \theta_{\bar{p}})}.
\label{eq:potential_of_p}
\end{align}
This potential reaches its minimum at the perihelion ($\theta_j=\theta_{\bar p}$) and maximum at the aphelion ($\theta_{\bar p} \pm \pi$). The resulting potential energy difference drives kinetic energy variations. Under the condition that the energy input from the natural frequency term balances dissipation, we obtain the following
\begin{align}
\mathcal{E}(\theta_j)= \mathcal{K}(\theta_j)+\mathcal{U}(\theta_j)
& = \dfrac{1}{2} m \dot{\theta_j}^2-KR_{P}\cos{(\theta_j - \theta_{\bar{p}})}  \cr
& = \text{constant.}
\end{align}
Our measurements of angular velocities $\dot \theta_{\bar{a}}$ ($a\in s,t,q$) as functions of the phase difference $\theta_{\bar{a}} -\theta_{\bar{p}}$ [Fig.~\ref{fig:fig_S_6}] reveal minimal velocities at the aphelion. These results demonstrate that HOCs formation is primarily governed by interactions with the SC rather than the PC. This leads to rational frequency relationships between HOC and SC angular velocities, thus establishing harmonic modes between these clusters.

\clearpage
\newpage


\section{\bfadhoc potential of HOC}
\label{sec:adhoc potential of HOC}

\begin{figure}[h]
\resizebox{1.0\columnwidth}{!}{\includegraphics{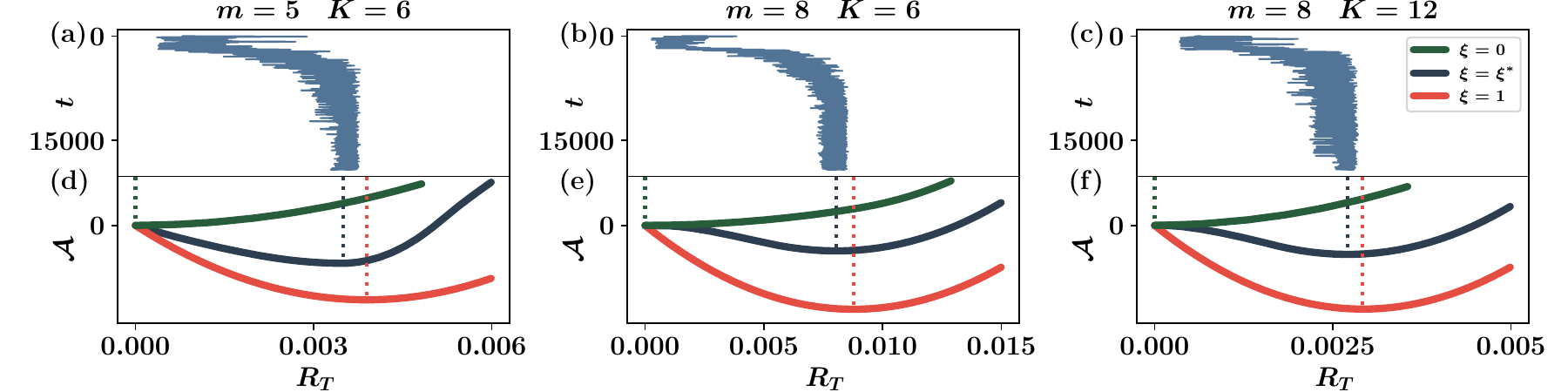}}
\caption{\textbf{Evolution of TC size and SC-HOC interaction effects.}
(a)--(c) Evolution of the TC order parameter $R_T$ versus time (increasing downward along $y$-axis) for different parameter sets ($m, K$). The curves converge to steady-state values $R_T(\infty)$.
(d)--(f) \adhoc potential $\mathcal{A}(R_T)$ versus $R_T$ calculated using three different weightings: $\xi=0$ (dark green), $\xi=\xi^*$ (dark blue), and $\xi=1$ (bright red). The orange curves, corresponding to the optimized weighting $\xi^*$, show best agreement with numerical results, validating the functional form proposed in Eq.~\eqref{eq:Work_Energy_H}.
The optimal weighting parameters are determined as $\xi^* = 0.072 \pm 0.030$ for $(m,K) = (5,6)$, $\xi^* = 0.036 \pm 0.018$ for $(m,K) = (8,6)$, and $\xi^* = 0.060 \pm 0.009$ for $(m,K) = (8,12)$. We observe that $\xi^*$ increases with decreasing $m$ or increasing $K$. This trend reflects stronger SC–HOC interactions near the aphelion, where lower oscillator velocities (due to decreased $m$ or increased $K$) result in extended interaction times.
To incorporate this SC coherence effect, we modified Eq.~\eqref{eq:Work_Energy_H} by replacing $K R_H$ with $K(\xi R_S+R_H)$ in the denominator, where $\xi$ represents the relative strength of the SC influence on $R_H$. The observation that $\xi^* < 1$ in all cases indicates a moderated SC effect compared to $\xi=1$.}
\label{fig:fig_S_7}	
\end{figure}

[Fig.~\ref{fig:fig_S_7}] illustrates the dynamics and stability of the TC for three parameter sets: $(m,K)$ = (5,6), (8,6), and (8,12). The temporal evolution of the TC order parameter $R_T$ and its corresponding \adhoc potential profiles validate our proposed functional form $f(R_T)$ in the self-consistency equation~\eqref{eq:Work_Energy_H}. Using this formulation, we construct the \adhoc potential $\mathcal{A}(R_T)$ defined in Eq.\eqref{eq:adhoc}, where the parameter $\xi$ is determined by matching the steady-state value $R_T(\infty)$ with the potential minimum $R_T^*$. The frequency boundaries of the TC are then obtained by numerically solving Eq.\eqref{eq:Work_Energy_H} using the fourth-order Runge-Kutta method.

\clearpage
\newpage


\section{The case of a large inertial: PC collapsing behavior}
\label{sec:when $m$ is large}

\begin{figure}[h]
\resizebox{1.0\linewidth}{!}{\includegraphics{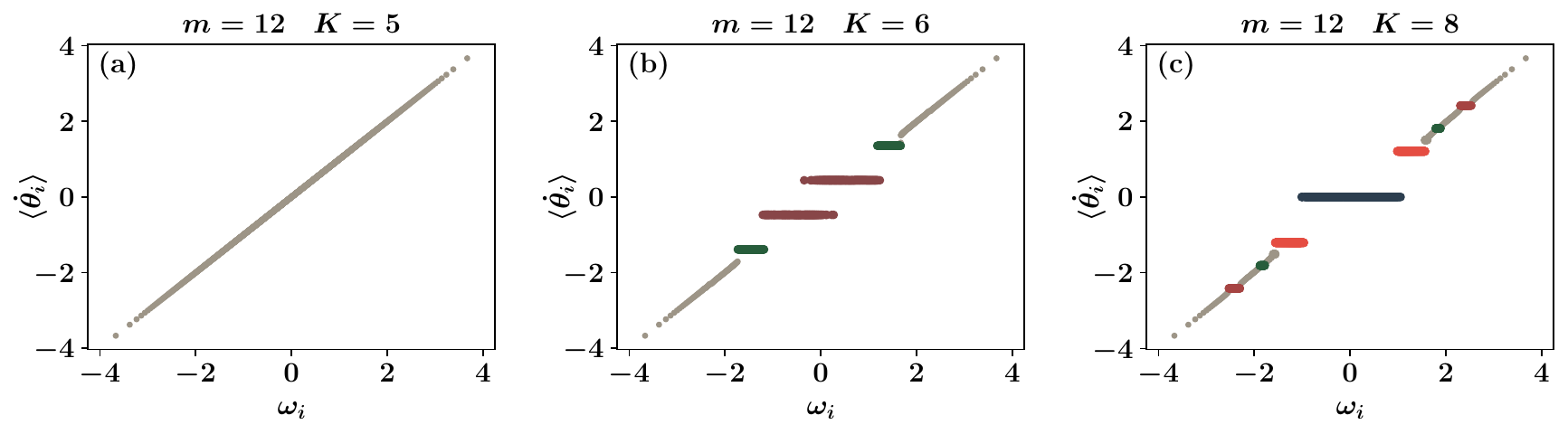}}
\caption{\textbf{PC collapsing behaviors.} (a) Oscillators are desynchronized when $K$ is small. (b) The initially formed PC collapses and disappears in the steady state when $K$ is moderate. (c) The PC is stable when $K$ is large.}
\label{fig:fig7}	
\end{figure}

For $m=12$, we identify three distinct regimes [Fig.~\ref{fig:fig7}]: no synchronization at weak coupling ($K < 4$), PC collapse with persistent SCs at intermediate coupling ($4 < K < 7$), and stable PC-SC coexistence at strong coupling ($K > 7$). The transitions between these regimes are characterized by sudden changes in the order parameters $R_p$ and $R_s$, with the intermediate regime showing particularly rich dynamics as oscillators escape from PC to SCs.

\clearpage
\newpage


\section{Analogous to the Shapiro steps in the shunted Josephson Junction array}
\label{sec:Analogous to the Shapiro steps in the shunted Josephson Junction array}

The dynamics of a resistor-capacitor shunted Josephson junction~\cite{strogatz2018nonlinear, josephson1962possible, shapiro1963josephson} can be described by the equation:
\begin{align}
C\dot{V}+\frac{V}{R}+I_c\sin\phi=I,
\label{eq:jj}
\end{align}
where the voltage difference $V=(\hbar/2e)\dot \phi$ and phase difference $\phi$ are measured across the junction. Here, $\hbar=h/(2\pi)$ represents the reduced Planck constant, $e$ is the elementary charge, $I_c$ denotes the critical current for the Josephson effect, and $I$ represents the total current in the system. This equation becomes equivalent to the second-order Kuramoto model when extended to an array of connected Josephson junctions. The corresponding I-V characteristics exhibit Shapiro steps with distinct plateaus~\cite{lee1990calculation, kvale1991theory, lee1991quantized, rzchowski1991frequency, yu1992resistance}, analogous to the Devil's staircase pattern described in the main text. While Shapiro's steps have been extensively investigated, previous studies primarily focused on two specific cases: the overdamped regime (small or zero $C$) and systems under external magnetic fields, where the sine term includes an additional magnetic flux component. For large capacitance values, research has been largely limited to hysteresis phenomena~\cite{yu1992resistance, dianorre2021dynamical, ngongiah2022resistive}. Thus, the dynamics we report here remain unexplored in the context of Josephson junction arrays with large capacitance, presenting an important direction for future investigation.


\clearpage
\newpage



\end{document}